\documentclass[11pt]{article}
\usepackage{amssymb,amsmath,verbatim,amsthm}
\usepackage{longtable}

\newtheorem{theorem}{Theorem}[section]
\newtheorem{lemma}[theorem]{Lemma}
\newtheorem{proposition}[theorem]{Proposition}
\newtheorem{corollary}[theorem]{Corollary}

\newtheorem{example}[theorem]{Example}
\newtheorem{remark}[theorem]{Remark}

\newtheorem{problem}[theorem]{Problem}

\newcommand{\ignore}[1]{}

\newcommand{\ord}{{\operatorname{ord}}}

\newcommand{\Tr}{{\operatorname{Tr}}}
\newcommand{\GF}{{\operatorname{GF}}}
\newcommand{\PG}{{\operatorname{PG}}}
\newcommand{\PGL}{{\operatorname{PGL}}}

\newcommand{\Supp}{{\operatorname{Supp}}}
\newcommand{\zero}{{\operatorname{zero}}}
\newcommand{\stab}{{\operatorname{Stab}}}

\newcommand{\Sx}{{\operatorname{S}}}
\newcommand{\wt}{{\operatorname{wt}}}
\begin{document}

\title{Infinite families of linear codes supporting more  $t$-designs\footnote{The research was supported by NSFC under Grant No. 11971053. }}
\author{
 {\small Qianqian Yan,}  {\small  Junling  Zhou}\\
 {\small Department of Mathematics}\\ {\small Beijing Jiaotong University}\\
  {\small Beijing  100044, China}\\
 {\small 19118010@bjtu.edu.cn}\\
{\small jlzhou@bjtu.edu.cn}\\
}
\date{ }
\maketitle

\begin{abstract}
Tang and Ding [IEEE IT 67 (2021) 244-254]
studied the class of narrow-sense BCH codes  $\mathcal{C}_{(q,q+1,4,1)}$ and their dual codes with $q=2^m$ and established that the codewords of the minimum (or the second minimum) weight in these codes support infinite families of 4-designs or 3-designs. Motivated by this, we further investigate the codewords of the next adjacent weight in such codes and discover more  infinite classes of $t$-designs with $t=3,4$. In particular, we prove that the codewords of weight $7$ in  $\mathcal{C}_{(q,q+1,4,1)}$ support $4$-designs when $m \geqslant 5$ is odd and $3$-designs when $m \geqslant 4$ is even, which provide infinite classes of simple $t$-designs with new parameters.
Another significant class of $t$-designs  we produce in this paper has supplementary designs with parameters
4-$(2^{2s+1}+ 1,5,5)$; these designs have the smallest  index among all the known simple 4-$(q+1,5,\lambda)$ designs derived from codes for prime powers $q$; and they are further
proved to be isomorphic to the 4-designs admitting the projective general linear group PGL$(2,2^{2s+1})$ as the automorphism group
constructed by Alltop in 1969.



\medskip\noindent{\bf 2000 MSC}: 05B05, 51E10, 94B15

\medskip\noindent {\bf Keywords}: BCH code, trace code, $t$-design, elementary symmetric polynomial, isomorphic.
\end{abstract}

\section{Introduction}
Let $\mathcal{P}$ be a $v$-element set  and let $\mathcal{B}$ be a collection of $k$-subsets of $\mathcal{P}$. If every $t$-subset of $\mathcal{P}$ is contained in exactly $\lambda$ elements of $\mathcal{B}$, then the incidence structure $\mathbb{D}=(\mathcal{P},\mathcal{B})$ is called a $t$-$(v, k, \lambda)$ {\em design}, or simply {\em $t$-design}. The elements of $\mathcal{P}$ and  $\mathcal{B}$ are called {\em points} and  {\em blocks}, and the parameters $t$ and $\lambda$ are called  {\em strength} and {\em index} respectively. Let $\binom{\mathcal{P}}{k}$ denote the set of all $k$-subsets of $\mathcal{P}$. Obviously, a $t$-design with $t=k$ or $k=v$ trivially exists; such  $t$-designs are called  {\em complete designs}. A $t$-design $(\mathcal{P},\mathcal{B})$ is said to be {\em simple} if $\mathcal{B}$ does not contain any repeated blocks. In this paper, we  will mainly consider simple $t$-designs and hence the expression ``$t$-design" will
mean  ``simple $t$-design", unless otherwise stated.  A  {\em Steiner system}, denoted by S$(t,k,v)$, is a $t$-$(v, k, \lambda)$ design where $t \geqslant 2$ and $\lambda=1$. It is clear that a $t$-$(v,k,\lambda)$ design  also forms an $s$-$(v,k,\lambda_{s})$ design with
\begin{equation}\label{1}
\lambda_{s}=\lambda\binom{v-s}{t-s} / \binom{k-s}{t-s}
\end{equation}
for any $s$ with $1 \leqslant s \leqslant t$.

Let  $\mathbb{D}=(\mathcal{P},\mathcal{B})$ be a $t$-$(v,k, \lambda)$ design. By \cite{2006Handbook1}, supp$(\mathbb{D}):= (\mathcal{P}, \{\mathcal{P}\setminus B: B \in \mathcal{B}\})$ is defined to be the {\em supplementary
design} of $\mathbb{D}$, which is a $t$-$(v, v-k, \lambda')$ design with
\begin{equation}\label{e2}
\lambda'=\lambda\binom{v-t}{k}/\binom{v-t}{k-t};
\end{equation}
and comp$(\mathbb{D}):=(\mathcal{P},\binom{\mathcal{P}}{k}\setminus \mathcal{B})$ is  defined to be  the {\em complementary design} of $\mathbb{D}$, which is a $t$-$(v, k, \lambda'')$ design with
\begin{equation}\label{e3}
\lambda''=\binom{v-t}{k-t}-\lambda.
\end{equation}
In what follows, the block set $\binom{\mathcal{P}}{k} \setminus \mathcal{B}$ is also denoted by $\overline{\mathcal{B}}$ for short.

Let $\mathcal{C}$ be an $[n,k,d]$ linear code over the finite field GF$(q)$. For $0\leqslant w \leqslant n$, denote by $A_w({\cal C})$ or  $A_{w}$ the number of codewords with Hamming weight $w$ in $\mathcal{C}$. The sequence $(A_{0},A_{1},\ldots, A_{n})$ is called the {\em weight distribution} of $\mathcal{C}$, and the polynomial $\sum^{n}_{w=0}A_{w}z^{w}$ is  the {\em weight enumerator} of $\mathcal{C}$. For convenience we index  the coordinates of all codewords in $\mathcal{C}$ by the elements in an $n$-set $\mathcal{P}(\mathcal{C})$.  For each $k$ with $A_{k}\neq 0$, always  let $\mathcal{B}_{k}(\mathcal{C})$ denote the set of the supports of all codewords in $\mathcal{C}$ with Hamming weight $k$.  If the pair $(\mathcal{P}(\mathcal{C}),\mathcal{B}_{k}(\mathcal{C}))$ forms a $t$-$(n,k,\lambda)$ design, then we say that the code $\mathcal{C}$ {\em holds} or {\em supports} a $t$-design. This is our main approach of  coding-theoretic construction for combinatorial designs in this paper.
There are other different means of generating $t$-designs.
  The Assmus-Mattson Theorem \cite{1969New1} is  very useful in producing $t$-designs from linear codes (see, for example, \cite{Cunsheng2018Design, 2006Handbook2}).  Tang et al. \cite{Tang2020Codes}  generalized Assmus-Mattson Theorem and their new criterion outperforms the original Assmus-Mattson Theorem in some special cases. Additionally, we have another fundamental method to  produce  $t$-designs by considering  codes   admitting  $t$-homogeneous or $t$-transitive automorphism groups (see, for example, \cite{Berger1999The,2020Infinite1,2003fun}).


An $[n,k,d]$ linear code $\mathcal{C}$ over $\GF(q)$ is  {\em cyclic} if any cyclic shift of a
codeword is again a codeword, i.e.,  $(c_{0},c_{1},\ldots, c_{n-1})\in \mathcal{C}$ implies $(c_{n-1},c_{0},\ldots,$\ $c_{n-2})\in \mathcal{C}$. We usually  treat codewords in a cyclic code as polynomials
in $\GF(q)[x]$.  This means if  $(c_{0},c_{1},\ldots, c_{n-1})\in \mathcal{C}$ then we associate the polynomial $c_{0}+c_{1}x+\cdots+c_{n-1}x^{n-1}\in \GF(q)[x]/(x^n-1).$ In such a way any cyclic  code of length $n$ over $\GF(q)$ corresponds to a subset of the residue class ring $\GF(q)[x]/(x^n-1)$. It is obvious that a linear code $\mathcal{C}$ is cyclic if and only if the
corresponding subset in $\GF(q)[x]/(x^n-1)$ is an ideal of the  $\GF(q)[x]/(x^n-1)$.

 Let $\mathcal{C}$ be a cyclic code of length $n$ over $\GF(q)$. Because $\GF(q)[x]/(x^n-1)$ is a principal ideal ring,  there is a unique monic polynomial $g(x)\in \GF(q)[x]$ of lowest degree such that $\mathcal{C}=\langle g(x)\rangle$. This polynomial $g(x)$ (as a divisor of $x^n-1$) is called the {\em generator polynomial} of $\mathcal{C}$.

Let $m= \ord_{n}(q)$ and let $\alpha$ be a generator of $\GF(q^m)^{*}:=\GF(q^m)\setminus \{0\}$. Let $\beta=\alpha^{(q^{m}-1)/n}$. Then clearly $\beta$ is a primitive $n$-th root of unity in $\GF(q^m)$. The {\em minimal polynomial} $\mathbb{M}_{\beta^{s}}(x)$ of $\beta^{s}$ over $\GF(q)$ is the monic polynomial of lowest degree over  $\GF(q)$ with $\beta^s$ as a zero.
Let $\delta,h$ be positive integers with $2\leqslant \delta \leqslant n$. A {\em BCH code} over GF$(q)$ with length $n$ and {\em designed distance} $\delta$, denoted by $\mathcal{C}_{(q,n,\delta,h)}$, is a cyclic code of length $n$ whose generator polynomial is
$$g(x)=\text{lcm}(\mathbb{M}_{\beta^{h}}(x),\mathbb{M}_{\beta^{h+1}}(x),\dots,\mathbb{M}_{\beta^{h+\delta-2}}(x))$$
with the least common multiple calculated over $\GF(q)$.
When $h=1$, the code $\mathcal{C}_{(q,n,\delta,h)}$ is called a {\em narrow-sense} BCH code.  As is known to us all, BCH codes are an important subclass of cyclic codes with  many attractive properties.

A code $\mathcal{C}$ is called a {\em maximum distance separable code} (or MDS code), if the minimum distance $d$ meets  $d=n-k+1$. The {\em Singleton defect} of an $[n,k,d]$ code $\mathcal{C}$ is defined to be def$(\mathcal{C}) = n-k+1-d$ and we say that $\mathcal{C}$ is an {\em A$^{s}$MDS code} if $s=$def$(\mathcal{C})$. Thus, A$^{0}$MDS code is the same as an MDS code and an A$^{1}$MDS code is also called an {\em almost MDS code} (AMDS code). Hence, AMDS codes have parameters $[n, k, n-k]$.
 A code $\mathcal{C}$ is said to be a {\em near MDS code} (NMDS code) if both $\mathcal{C}$ and $\mathcal{C}^{\bot}$ are AMDS codes.

 MDS codes do hold $t$-designs with large $t$, but these designs are all trivial unfortunately.  Ding and Tang \cite{Cunsheng2020Infinite} presented an infinite family of NMDS codes over $\GF(3^m)$ holding  $3$-designs and an infinite family of NMDS codes over $\GF(2^{2s})$ holding $2$-designs. The first  NMDS code supporting a $4$-design was the $[11,6,5]$ ternary Golay code which  holds a Steiner system S$(4,5,11)$ discovered in 1949 by Golay \cite{Golay1949Notes}. Recently Tang and Ding \cite{Tang2020An}
studied  the narrow-sense BCH codes $\mathcal{C}_{(q,q+1,4,1)}$ and the dual codes $\mathcal{C}^\bot_{(q,q+1,4,1)}$ where $q=2^m$. It was shown that the codewords of the
minimum (or the second minimum) weight  in $\mathcal{C}_{(q,q+1,4,1)}$ support $4$-designs when $m \geqslant 5$ is odd and they support $3$-designs when $m \geqslant 4$ is even. In this paper, we will further investigate the codewords of the next adjacent weight (of weight 7) and prove that ${\cal B}_7(\mathcal{C}_{(q,q+1,4,1)})$ also supports a $4$-design when $m \geqslant 5$ is odd and it supports a $3$-design when $m \geqslant 4$ is even.  For $q=2^m$ with odd $m\geqslant 5$, the minimum weight codewords in   $\mathcal{C}^\bot_{(q,q+1,4,1)}$  were shown to support  $4$-$(q+1,q-5,\lambda)$ designs in \cite{Tang2020An} and in this paper we will show the second minimum weight codewords (of weight $q-4$) support   again  4-designs. It is very interesting and significant that the supplementary designs of these 4-designs have parameters $4$-$(q+1,5,5)$,  because on one hand they achieve the smallest    index among all the known simple $4$-$(q+1,5,\lambda)$ designs (with $q$ prime powers) derived from codes and on the other hand they are proved to be isomorphic to the 4-designs   admitting the projective general linear group PGL$(2,q)$ as the automorphism group constructed by Alltop \cite{1969An} in 1969. 
This paper is not only a sequel of  \cite{Tang2020An},  but also we define  generalized combinatorial objects, adopt new approaches and produce
 more infinite families of $t$-designs with new parameters.


 The rest of the paper is organized as follows. In Section 2, we introduce the definition of elementary symmetric polynomials (ESPs) and  develop block sets produced from three types of variants of ESPs. We also document several useful results produced in  \cite{Tang2020An}. In Section 3, we present several new infinite families of $t$-designs with $t=3, 4$ by exploring the block sets generated from the variants of ESPs. In Section 4, we consider the narrow-sense BCH codes $\mathcal{C}_{(q,q+1,4,1)}$  with $q=2^m$ and  prove that the codewords of weight $7$ in $\mathcal{C}_{(q,q+1,4,1)}$ support $4$-designs when $m \geqslant 5$ is odd and they support $3$-designs when $m \geqslant 4$ is even. In Section 5, we represent the dual codes $\mathcal{C}^\bot_{(q,q+1,4,1)}$ ($q=2^m,m \geqslant 5$  odd)  in terms of trace codes in order to study the automorphism groups of their support sets. Then we prove that  the codewords of weight $q-4$ support 4-designs, whose supplementary designs are isomorphic to the  $4$-$(q+1,5,5)$ designs constructed by Alltop \cite{1969An}. In Section 6 we summarize our main results  and  conclude the paper with some remarks.

\section{Preliminaries}

A polynomial $f$ is  {\em symmetric} if the polynomial is invariant under all permutations of its variables. The {\em elementary symmetric polynomial} (briefly by ESP) of degree $l$ in $k$ variables $u_{1},u_{2},\ldots,u_{k}$, is defined by
$$\sigma_{k,l}(u_{1},u_{2},\ldots,u_{k})=\sum \limits_{I\subseteq [k] \atop{| I|=l}}\prod\limits_{j\in I}u_{j}.$$
For simplicity, $\sigma_{k,l}(u_{1},u_{2},\ldots,u_{k})$ is also denoted by $\sigma_{k,l}(B)$ where $B=\{u_{1},u_{2},\ldots,u_{k}\}$ or briefly denoted $\sigma_{k,l}$ if the context is clear.

Throughout the paper we always let $q$ be a prime power and let $U_{q+1}$ be the set of all $(q+1)$-th roots of unity in $\GF(q^2)$, that is,
\begin{equation*}
U_{q+1} = \{u \in \text{GF}(q^{2}): u^{q+1} = 1\}.
\end{equation*}
Let $f$ be a  symmetric polynomial in $k$ variables whose coefficients are taken in GF$(q^2)$. We define
\begin{equation}\label{Bf}
\mathcal{B}_{f,q+1}=\bigg\{\{u_{1},u_{2},\ldots ,u_{k}\}\in \binom{U_{q+1}}{k}:f(u_{1},u_{2},\ldots,u_{k})=0\bigg\}.
\end{equation}
This produces an incidence structure $\mathbb{D}=(U_{q+1},\mathcal{B}_{f,q+1})$. If $\mathbb{D}$ forms a $t$-$(q+1,k,\lambda)$ design, then we say that  $f$ supports a $t$-$(q+1,k, \lambda)$ design. In particular, for any positive integer $k\leqslant q+1$, the block set $\mathcal{B}_{\sigma_{k,l},q+1}$ produced from the ESP $\sigma_{k,l}$ is defined by
\begin{equation}\label{B}
\mathcal{B}_{\sigma_{k,l},q+1}=\bigg \{B\in \binom{U_{q+1}}{k}  : \sigma _{k,l}(B)=0 \bigg \}.
\end{equation}

In \rm{\cite{Tang2020An}}, Tang and Ding presented several infinite families of linear codes supporting $t$-designs, whose block sets are isomorphic to $\mathcal{B}_{\sigma_{k,l},q+1}$ where $(k,l,q)\in \{(4,2,2^{2s}), (4,2,3^m), (5,2,2^{2s}),$ \ $(6,3,2^m)\}$. In this paper we also concentrate on the topic of linear codes supporting $t$-designs by handling the incidence structures produced primarily from three types of variants of ESPs.

We define a block set $\mathcal{B}^{u}_{\sigma_{k,l},q+1}$ by
\begin{equation}\label{Bu}
\mathcal{B}^{u}_{\sigma_{k,l},q+1}=\bigg \{B\in \binom{U_{q+1}}{k} : \sigma _{k,l}(B-a)=0 \ \text{for some}\ a \in U_{q+1}\bigg \},
\end{equation}
where $B-a:=\{b-a: b \in B\}$.
If $(U_{q+1},\mathcal{B}^{u}_{\sigma_{k,l},q+1})$ forms a $t$-$(q+1,k,\lambda)$ design, we say that the {\em $u$-variant} of the ESP $\sigma_{k,l}$ supports a $t$-$(q+1,k, \lambda)$ design.

More restrictively, we define the block sets $\mathcal{B}_{\sigma_{k,l},q+1}^{b}$ and  $\mathcal{B}_{\sigma_{k,l},q+1}^{\overline{b}}$ generated from the {\em $b$-variant} and {\em $\overline{b}$-variant} of ESPs $\sigma_{k,l}$ respectively by
\begin{equation}\label{Bv}
\mathcal{B}^{b}_{\sigma_{k,l},q+1}=\bigg\{B\in \binom{U_{q+1}}{k}: \sigma _{k,l}(B-a)=0 \ \text{for some}\ a \in B\bigg\},
\end{equation}
and
\begin{equation}\label{Bw}
\mathcal{B}^{\overline{b}}_{\sigma_{k,l},q+1}=\mathcal{B}^{u}_{\sigma_{k,l},q+1} \setminus \mathcal{B}^{b}_{\sigma_{k,l},q+1}.
\end{equation}
%

Let $B=\{u_{1},u_{2},\dots,u_{k}\}\in \binom{U_{q+1}}{k}$. One gets
\begin{equation}\label{ba}
\sigma_{k,l}(B-a)=\sum \limits_{I\subseteq [k] \atop {|I|=l}}\prod\limits_{j\in I} (u_{j}-a)=\sum\limits^{l}_{i=0}(-a)^{l-i}\binom{k-i}{l-i}\sigma_{k,i}(B).
\end{equation}
Then $B \in \mathcal{B}^{u}_{\sigma_{k,l},q+1}$ if and only if $\sum\limits^{l}_{i=0}(-a)^{l-i}\binom{k-i}{l-i}\sigma_{k,i}(B)=0$ for some $a \in U_{q+1}$.

In this paper we will present new infinite families of $t$-designs with $t=3,4$ by exploring the block sets generated from the $u$-variant, $b$-variant, or $\overline{b}$-variant of ESPs. The effect is twofold. From the perspective of coding theory, the three types of variants usually provide succinct  descriptions for the designs held by the codewords of a fixed weight in a linear code.  On the other hand, from the aspect of design theory, we produce several infinite families of $t$-designs. Comparing with \cite{2006Handbook1}, we find that most of these parameters are new and  an attractive feature of these designs lies in the simpleness of the designs although the indices $\lambda$ are somewhat large.


Now we record several conclusions in \cite{Tang2020An} and derive some corollaries for later use.

\begin{lemma} {\rm\cite[Lemmas 20, 22, 24-26]{Tang2020An}}\label{lemma11} Let $q=2^{m}$ and $m\geqslant 4$.
Let $\{u_{1},u_{2},u_{3},u_{4}\} \in \binom{U_{q+1}}{4}$ such that $\sigma_{5,2}(u_{1},u_{2},u_{3},u_{4},
u_{5}) \neq 0$ for any $u_{5} \in U_{q+1} \setminus \{u_{1},u_{2},u_{3},u_{4}\}$. Define $S_{1}$ and $S$ by
\begin{equation}\label{S1}
S_{1}=\bigg\{\frac{\sigma_{4,3}+u_{i}\sigma_{4,2}}{\sigma_{4,2}+u_{i}\sigma_{4,1}}:i=1,2,3,4 \bigg\}  \bigcup \bigg\{\sqrt{\frac{\sigma_{4,3}}{\sigma_{4,1}}} \bigg\}
\end{equation}
and
\begin{equation}\label{S}
S=S_{1}\cup \{u_{i}:i=1,2,3,4\}.
\end{equation}
Then we have the following.
\begin{enumerate}
\item[$(1)$] $S\subseteq U_{q+1}$, $|S_{1}|=5$ and $|S|=9$.
  \item[$(2)$] If $m$ is even and $\{\alpha,\beta\}\in \binom{U_{q+1}}{2}$ such that $\sigma_{5,2}(u_{1},u_{2},u_{3},\alpha,\beta)=0$, then $\alpha,\beta\notin S$.
\item[$(3)$] Let $u=\frac{\sigma_{5,3}(u_{1},u_{2},u_{3},u_{4},u_{5})}{\sigma_{5,2}(u_{1},u_{2},u_{3},u_{4},u_{5})}$. Then $u\in \{u_{1},u_{2},u_{3},u_{4},u_{5}\}$ if and only if $u_{5}\in S_{1}$.
\end{enumerate}
\end{lemma}

\begin{lemma}{\rm\cite[Lemma 17]{Tang2020An}}\label{lemma22}
For $q=2^{m}$ with odd $m\geqslant 5$  and $B\in \binom{U_{q+1}}{5}$, one has $\sigma_{5,2}(B)\neq 0$.
\end{lemma}

\begin{lemma}{\rm\cite[Theorem 3]{Tang2020An}}\label{lemma23}
For $q = 2^{m}$ where $m\geqslant 4$ is even,  $(U_{q+1}, \mathcal{B}_{\sigma_{5,2},q+1})$  forms a Steiner
system $\Sx (3,5,q+1)$.
\end{lemma}
%
%

\begin{lemma}{\rm\cite[Theorem 2]{Tang2020An}}\label{lemma 24}
For $q = 2^{m}$ where $m\geqslant 5$ is odd,  $(U_{q+1}, \mathcal{B}_{\sigma_{6,3},q+1})$ forms a $4$-$(q+1,6,\frac{q-8}{2})$ design.
\end{lemma}

\proof Let $\{u_{1},u_{2},u_{3},u_{4}\}$ be a fixed $4$-subset of $U_{q+1}$. For any $u_{5}\in U_{q+1} \setminus \{u_{1},u_{2},u_{3},u_{4}\}$, $\sigma_{5,2}(u_{1},u_{2},u_{3},u_{4},u_{5})\ne 0$ from Lemma \ref{lemma22}. Define
$$\mathcal{T}=\bigg\{\{u_{1},u_{2},u_{3},u_{4},u_{5},u_{6}\}\in \binom{U_{q+1}}{6}: u_{5}\in U_{q+1}\setminus S, u_{6}=\frac{\sigma_{5,3}(u_{1},u_{2},u_{3},u_{4},u_{5})}{\sigma_{5,2}(u_{1},u_{2},u_{3},u_{4},u_{5})}\bigg\},$$
where $S$ is given by Eq. (\ref{S}). Note that $\sigma_{5,3}=\sigma_{5,5}\sigma_{5,2}^{q}$. Then $(\frac{\sigma_{5,3}}{\sigma_{5,2}})^{q+1}=\sigma_{5,5}^{q+1}(\sigma_{5,2}^{q-1})^{q+1}=1$. This shows that $\frac{\sigma_{5,3}}{\sigma_{5,2}} \in U_{q+1}$. It is easily checked that $E\subseteq B$ and $B\in \mathcal{B}_{\sigma_{6,3},q+1}$ if and only if $B\in \mathcal{T}$. From Lemma \ref{lemma11} (3), $\frac{\sigma_{5,3}}{\sigma_{5,2}}\not\in S$ if $u_{5}\not\in S$. Noting the symmetry of $u_{5}$ and $u_{6}$, we have $|\mathcal{T}|=\frac{q+1-9}{2}$ by Lemma \ref{lemma11} (1). As a result, $(U_{q+1}, \mathcal{B}_{\sigma_{6,3},q+1})$ is a $4$-$(q+1,6,\frac{q-8}{2})$ design. \qed

\begin{lemma}{\rm\cite[Theorem 4]{Tang2020An}}\label{l25}
For $q = 2^{m}$ with even $m\geqslant 4$,  $(U_{q+1}, \mathcal{B}_{\sigma_{6,3},q+1})$ gives a $3$-$
(q+1,6,\frac{(q-4)^2}{6})$ design.
\end{lemma}

\proof Let $E=\{u_{1},u_{2},u_{3}\}$ be a fixed 3-subset of $U_{q+1}$. By Lemma \ref{lemma23}, there is a unique block $A \in \binom{U_{q+1}}{5}$ such that $E\subseteq A$ and $\sigma_{5,2}(A)=0$. Set
$$\mathcal{T}_{1}=\{A\cup \{u_{i}\}: u_{i} \in U_{q+1}\setminus A\},$$
and
$$\mathcal{T}_{2}=\bigg\{\left.
                         \begin{array}{c}
                           \{u_{1},u_{2},u_{3},u_{4},u_{5},u_{6}\}\in \binom{U_{q+1}}{6}: \\
                            \\
                         \end{array}
                       \right.
 \left.
                                                                  \begin{array}{l}
 u_{4}\in U_{q+1}\setminus A, u_{5}\in U_{q+1}\setminus (A\cup S), \\
 u_{6}=\frac{\sigma_{5,3}(u_{1},u_{2},u_{3},u_{4},u_{5})}{\sigma_{5,2}(u_{1},u_{2},u_{3},u_{4},u_{5})}  \\
                                                                  \end{array}
                                                                \right.
\bigg\},$$
where $S$ is given by (\ref{S}). Let $\mathcal{T}=\mathcal{T}_{1}\cup\mathcal{T}_{2}$. Obviously $\mathcal{T}_{1}$ and $\mathcal{T}_{2}$ are disjoint. It is easily checked that $E\subseteq B$ and $B\in \mathcal{B}_{\sigma_{6,3},q+1}$ if and only if $B\in \mathcal{T}$. Note that $|\mathcal{T}_{1}|=q+1-5=q-4$ and $|\mathcal{T}_{2}|=\frac{(q+1-|A|)(q+1-|A\cup S|)}{6}$. By Lemma \ref{lemma11}, $|A\cup S|=11$. Then $|\mathcal{T}_{2}|=\frac{(q-4)(q-10)}{6}$ and $(U_{q+1}, \mathcal{B}_{\sigma_{6,3},q+1})$ is a $3$-$(q+1,6,\frac{(q-4)^2}{6})$ design. \qed

The following corollary follows from the proof of Lemma \ref{l25}.

\begin{corollary}\label{c26}
Let $q = 2^{m}$ with even  $m\geqslant 4$. Let $E$ be a fixed $3$-subset of $U_{q+1}$.
\begin{enumerate}
  \item[$(1)$] Set $\mathcal{T}_{1}=\{B\in \mathcal{B}_{\sigma_{6,3},q+1}: E \subseteq B, B \setminus \{u\} \in \mathcal{B}_{\sigma_{5,2},q+1}\ \text{for some}\ u\in B\setminus E \}$. Then $|\mathcal{T}_{1}| = q-4$.
\item[$(2)$] Set $\mathcal{T}_{2}=\{B\in \mathcal{B}_{\sigma_{6,3},q+1}: E \subseteq B, B \setminus \{u\} \notin \mathcal{B}_{\sigma_{5,2},q+1}\  \text{for any} \ u\in B\setminus E \}$. Then $|\mathcal{T}_{2} |= \frac{(q-4)(q-10)}{6}$.
\end{enumerate}
\end{corollary}

Combining Lemma \ref{lemma23} with the proofs of Lemmas \ref{lemma 24} and \ref{l25} gives the following corollary, which is very useful in this paper.

\begin{corollary}\label{c27}
\begin{enumerate}
\item[$(1)$] Let $q = 2^{m}$ with even $m\geqslant 4$. For any $B_{1}, B_{2}\in \mathcal{B}_{\sigma_{5,2},q+1}$ with $B_{1} \neq B_{2}$, we have $|B_{1}\cap B_{2}|\leqslant 2$.
\item[$(2)$] Let $q = 2^{m}$ with even $m\geqslant 4$. For any $B_{1}, B_{2}\in \mathcal{B}_{\sigma_{6,3},q+1}$ with $B_{1} \neq B_{2}$, we have $|B_{1}\cap B_{2}|\leqslant 5$; equality occurs only if $\sigma_{5,2}(B_{1} \cap B_{2})=0$.
\item[$(3)$] Let $q = 2^{m}$ with odd $m \geqslant 5$. For any $B_{1}, B_{2}\in \mathcal{B}_{\sigma_{6,3},q+1}$ with $B_{1} \neq B_{2}$, we have $|B_{1}\cap B_{2}|\leqslant 4$.
\end{enumerate}
\end{corollary}

\begin{lemma}{\rm\cite[Theorem 4]{Tang2020An}}\label{l10}
Let $q = 2^{m}$ with even $m \geqslant 4$. Then  $(U_{q+1}, \mathcal{B}^{0}_{\sigma_{6,3},q+1})$ forms a $3$-$(q+1,6,2(q-4))$ design and  $(U_{q+1}, \mathcal{B}_{\sigma_{6,3},q+1} \setminus \mathcal{B}^{0}_{\sigma_{6,3},q+1})$ gives a $3$-$(q+1,6,\frac{(q-4)(q-16)}{6})$ design, where the block set $\mathcal{B}^{0}_{\sigma_{6,3},q+1}$ is defined by
\begin{equation}\label{0}
\mathcal{B}^{0}_{\sigma_{6,3},q+1}=\bigg\{B\in \binom{U_{q+1}}{6}: \sigma_{5,2}(B\setminus \{u\})=0\ \text{for some}\ u\in B\bigg\}.\end{equation}
\end{lemma}

To conclude this section we give two theorems showing several cases when the ESP $\sigma_{k,l}$ and the $u$-variant of $\sigma_{k,l}$ support the same designs.

\begin{theorem}\label{t28}
For $q=2^{m}$ with even $m \geqslant 4$,  $(U_{q+1},\mathcal{B}^{u}_{\sigma_{5,2},q+1})$ forms a Steiner system $\Sx (3,5,q+1)$.
\end{theorem}

\proof Let $B\in \binom{U_{q+1}}{5}$. By Eq. (\ref{Bu}), $B \in \mathcal{B}^{u}_{\sigma_{5,2},q+1}$ if and only if $\sigma_{5,2}(B-a)=0$ for some $a\in U_{q+1}$. From Eq. (\ref{ba}), $\sigma_{5,2}(B-a)=\sigma_{5,2}-4a\sigma_{5,1}+10a^{2}$. Since $q=2^m$, $\sigma_{5,2}(B-a)=0$ is the same as $\sigma_{5,2}(B)=0$. This shows that $\mathcal{B}^{u}_{\sigma_{5,2},q+1}=\mathcal{B}_{\sigma_{5,2},q+1}$.  So the conclusion
follows by Lemma \ref{lemma23}.\qed

\begin{theorem}\label{t29}
\begin{enumerate}
\item[$(1)$] For $q=2^{m}$ with even $m \geqslant 4$,  $(U_{q+1},\mathcal{B}^{u}_{\sigma_{6,3},q+1})$ forms a $3$-$
(q+1,6,\frac{(q-4)^2}{6})$ design.
\item[$(2)$] Let $q=2^{m}$ with odd $m \geqslant 5$, $(U_{q+1},\mathcal{B}^{u}_{\sigma_{6,3},q+1})$ forms a $4$-$(q+1,6,\frac{q-8}{2})$ design.
\end{enumerate}
\end{theorem}

\proof Let $B\in \binom{U_{q+1}}{6}$. By Eq. (\ref{Bu}), $B \in \mathcal{B}^{u}_{\sigma_{6,3},q+1}$ if and only if $\sigma_{6,3}(B-a)=0$ for some $a\in U_{q+1}$. From Eq. (\ref{ba}),
$ \sigma_{6,3}(B-a)=\sigma_{6,3}-4a\sigma_{6,2}+10a^{2}\sigma_{6,1}-20a^{3}$. Because $q=2^m$, $\sigma_{6,3}(B-a)=0$ is the same as $\sigma_{6,3}(B)=0$. This shows that $\mathcal{B}^{u}_{\sigma_{6,3},q+1}=\mathcal{B}_{\sigma_{6,3},q+1}$ and the conclusion can be drawn from Lemmas \ref{lemma 24} and \ref{l25}.  \qed

\section{$t$-Designs from variants of ESPs}
In the rest of the whole paper we always let $q=2^m$ with $m\geqslant 4$ being an integer. The goal of this section is to construct new infinite families $t$-designs for $t=3,4$ from three types of variants of ESPs defined  in Section 2. These results would play an important part in proving that the BCH codes $\mathcal{C}_{(q,q+1,4,1)}$ and their dual  $\mathcal{C}^{\bot}_{(q,q+1,4,1)}$ support $4$-designs or $3$-designs, see for details in the next two sections.

 We stipulate a notation for this section. For two sets $A$ and $B$, we use $A\uplus B$ to denote the multiset union of $A$ and $B$ while $A\cup B$ denotes their ordinary union.

\begin{theorem}\label{t3.2}
For $q=2^{m}$ with even $m \geqslant 4$, $(U_{q+1},\mathcal{B}^{u}_{\sigma_{4,2},q+1})$ forms a $3$-$(q+1,4,2)$ design.
\end{theorem}

\proof By Eq. (\ref{Bu}), $B \in \mathcal{B}^{u}_{\sigma_{4,2},q+1}$ if and only if $B\in \binom{U_{q+1}}{4}$ and $\sigma_{4,2}(B-a)=0$ for some $a\in U_{q+1}$. Let $B=\{u_{1},u_{2},u_{3},u_{4}\}$ be any block of $\mathcal{B}_{\sigma_{4,2},q+1}^{u}$. From Eq. $(\ref{ba})$ one has $$\sigma_{4,2}(B-a)=\sigma_{4,2}(B)-3a\sigma_{4,1}(B)+6a^{2}=\sigma_{4,2}(B)+a\sigma_{4,1}(B).$$
Then $\sigma_{4,2}(B-a)=0$  is equivalent to $\sigma_{5,2}(B\uplus\{a\})=0$. Next we prove $a \notin B$. Otherwise $a \in B$. W.l.o.g. let $a=u_{4}$ and $\sigma_{4,2}+u_{4}\sigma_{4,1}=0$. Thus
$$\sigma_{3,2}+u_{4}^{2}= (\sigma_{3,2}+u_{4}\sigma_{3,1})+u_{4}(\sigma_{3,1}+u_{4})=\sigma_{4,2}+u_{4}\sigma_{4,1}=0.$$
Hence, we have
\begin{eqnarray*}&&u_{4}^{2}\sigma_{3,1}+\sigma_{3,3} =u_{4}^{2}(\sigma_{3,3}\sigma_{3,2}^{q})+\sigma_{3,3} =u_{4}^{2}\sigma_{3,3}(\sigma_{3,2}^{q}+u_{4}^{-2}) =u_{4}^{2}\sigma_{3,3}(\sigma_{3,2}^{q}+u_{4}^{2q})\\  && \qquad  \qquad \quad \ \ =u_{4}^{2}\sigma_{3,3}(\sigma_{3,2}+u_{4}^{2})^{q} =0.\end{eqnarray*}
Multiplying both sides of $\sigma_{3,2}+u_{4}^{2}=0$ by $u_{4}$ and then adding $u_{4}^{2}\sigma_{3,1}+\sigma_{3,3}$ yields
$$\sigma_{3,3}+u_{4}\sigma_{3,2}+u_{4}^{2}\sigma_{3,1}+u_{4}^{3}=0.$$
Then
$$(u_{4}+u_{1})(u_{4}+u_{2})(u_{4}+u_{3})=0,$$
which is contrary to the assumption $\{u_{1},u_{2},u_{3},u_{4}\} \in \binom{U_{q+1}}{4}$. It follows that $a\not \in B$ and thus $$\mathcal{B}_{\sigma_{4,2},q+1}^{u}=\Bigg\{B\in \binom{U_{q+1}}{4}: \sigma_{5,2}(B\cup \{a\})=0\ \text{for some}\ a\in U_{q+1}\setminus B\Bigg\}.$$

 Let $E=\{u_{1},u_{2},u_{3}\}$ be a fixed 3-subset of $U_{q+1}$. From Lemma \ref{lemma23}, $(U_{q+1},\mathcal{B}_{\sigma_{5,2},q+1})$ is a $3$-$(q+1,5,1)$ design. Then there is a unique pair $\{\alpha,\beta\} \in\binom{ U_{q+1}\setminus E}{2}$ such that $E\cup \{\alpha,\beta\}\in \mathcal{B}_{\sigma_{5,2},q+1}$ and so $E\cup \{\alpha\}$, $E\cup\{\beta\} \in \mathcal{B}^{u}_{\sigma_{4,2},q+1}$. It readily yields that $(U_{q+1},\mathcal{B}^{u}_{\sigma_{4,2},q+1})$ forms a  $3$-$(q+1,4,2)$ design.  \qed

\begin{lemma}\label{l3.3}
Let $q=2^{m}$ with $m \geqslant 4$. Let $B\in \binom{U_{q+1}}{5}$ such that $\sigma_{5,2}(B)= 0$. Then $\sigma_{6,3}(B\uplus \{u\})=0$ for any $u\in U_{q+1}$.
\end{lemma}

\proof Since $\sigma_{5,3}(B)=\sigma_{5,5}(B)\sigma_{5,2}^{q}(B)$, then $\sigma_{5,3}(B)=0$ if and only if $\sigma_{5,2}(B)=0$ by noting $\sigma_{5,5}(B)\ne 0$. Note that
$$\sigma_{6,3}(B\uplus\{u\})= \sigma_{5,3}(B)+u\sigma_{5,2}(B).$$
Then $\sigma_{6,3}(B\uplus\{u\})=0$ if $\sigma_{5,2}(B)= 0$. \qed

\begin{theorem}\label{t3.4}
For $q=2^{m}$ with odd $m \geqslant 5$, $(U_{q+1},\mathcal{B}_{\sigma_{5,3},q+1}^{u})$ is a complete $4$-design. In particular we have the following.
\begin{enumerate}
  \item[$(1)$]  $(U_{q+1},\mathcal{B}^{b}_{\sigma_{5,3},q+1})$ forms a $4$-$(q+1,5,5)$ design.
  \item[$(2)$]  $(U_{q+1},\mathcal{B}^{\overline{b}}_{\sigma_{5,3},q+1})$ forms a $4$-$(q+1,5,q-8)$ design.
\end{enumerate}
\end{theorem}

\proof Let $B\in \binom{U_{q+1}}{5}$. By Eq. (\ref{Bu}), $B \in \mathcal{B}^{u}_{\sigma_{5,3},q+1}$ if and only if $\sigma_{5,3}(B-a)=0$ for some $a\in U_{q+1}$. From Eq. (\ref{ba}) one obtains $$\sigma_{5,3}(B-a)=\sigma_{5,3}(B)-3a\sigma_{5,2}(B)+6a^{2}\sigma_{5,1}(B)-10a^{3}=\sigma_{5,3}(B)+a\sigma_{5,2}(B).$$
Then $\sigma_{5,3}(B-a)=0$ is equivalent to $\sigma_{6,3}(B\uplus \{a\})=0$. Let $E=\{u_{1},u_{2},u_{3},u_{4}\}$ be a fixed $4$-subset of $U_{q+1}$ and $B=\{ u_{1},u_{2},u_{3},u_{4},u_{5}\}$ be any block of $\mathcal{B}_{\sigma_{5,3},q+1}^{u}$ containing $E$. Then we consider the following two cases.
\begin{enumerate}
  \item[$(1)$] $B\in \mathcal{B}^{b}_{\sigma_{5,3},q+1}$. Then there is $u_{i}\in B$  such that $\sigma_{6,3}(B\uplus \{u_{i}\})=0$ by (\ref{Bv}). From Lemma \ref{lemma22}, $\sigma_{5,2}(B) \neq 0$. Note that
$$\left.
    \begin{array}{cl}
      \sigma_{6,3}(B\uplus\{u_{i}\}) &= \sigma_{5,3}(B)+u_{i}\sigma_{5,2}(B). \\
    \end{array}
  \right.$$
Then $u_{i}=\frac{\sigma_{5,3}(B)}{\sigma_{5,2}(B)}$. From Lemma \ref{lemma11} (3), $\frac{\sigma_{5,3}(B)}{\sigma_{5,2}(B)}\in B$ if and only if $u_{5}\in S_{1}$, where $S_{1}$ is given by Eq. (\ref{S1}). Hence, $|\{B\in \mathcal{B}_{\sigma_{5,3},q+1}^{b}: E\subseteq B\}|=|S_{1}|=5$ by Lemma \ref{lemma11} (1). So $E$ is contained in five distinct blocks of $\mathcal{B}^{b}_{\sigma_{5,3},q+1}$ and the incidence structure $(U_{q+1},\mathcal{B}^{b}_{\sigma_{5,3},q+1})$ is a  $4$-$(q+1,5,5)$ design.
\item[$(2)$] $B\in \mathcal{B}^{\overline{b}}_{\sigma_{5,3},q+1}$. Then there is $a \in U_{q+1} \setminus B$ such that $\sigma_{6,3}(B\cup \{a\})=0$ by (\ref{Bw}). From Lemma \ref{lemma 24}, $(U_{q+1},\mathcal{B}_{\sigma_{6,3},q+1})$ is a $4$-$(q+1,6,\frac{q-8}{2})$ design. Then there are $\frac{q-8}{2}$ blocks of $\mathcal{B}_{\sigma_{6,3},q+1}$ containing $E$. Noticing each block of $\mathcal{B}_{\sigma_{6,3},q+1}$ containing $E$ provides two blocks of $\mathcal{B}_{\sigma_{5,3},q+1}^{\overline{b}}$ containing $E$ and applying Corollary \ref{c27} (3), we know that $(U_{q+1},\mathcal{B}^{\overline{b}}_{\sigma_{5,3},q+1})$ is a $4$-$(q+1,5,q-8)$ design.
\end{enumerate}
Combining (1) and (2) yields that $(U_{q+1},\mathcal{B}_{\sigma_{5,3},q+1}^{u})$ is a $4$-$(q+1,5,q-3)$ design. This is a complete $4$-design. \qed

\begin{remark} Let $q=2^{m}$ and $m \geqslant 4$ be even.
 If we take each block $B\in \mathcal{B}^{b}_{\sigma_{5,3},q+1}$ such that $\sigma_{5,2}(B)=0$ five times and each of the other blocks of $\mathcal{B}^{b}_{\sigma_{5,3},q+1}$ once, then we produce a non-simple $4$-$(q+1,5,5)$ design by applying similar arguments to the proof of Theorem $\ref{t3.4}$ $(1)$.
\end{remark}

\begin{theorem}\label{t3.5}
For $q=2^{m}$ with even $m \geqslant 4$,  $(U_{q+1},\mathcal{B}^{\overline{b}}_{\sigma_{5,3},q+1})$ forms a $3$-$(q+1,5,\frac{q^2-10q+26}{2})$ design.
\end{theorem}

\proof Let $B\in \binom{U_{q+1}}{5}$. Analogous  to the proof of Theorem \ref{t3.4}, we have $B \in \mathcal{B}^{\bar{b}}_{\sigma_{5,3},q+1}$ if and only if $\sigma_{6,3}(B\cup \{a\})=0$ for some $a\in U_{q+1}\setminus B$ by noting Eq. (\ref{Bw}). Let $E=\{u_{1},u_{2},u_{3}\}$ be a fixed $3$-subset of $U_{q+1}$ and let $B$ be any block of $\mathcal{B}_{\sigma_{5,3},q+1}^{\overline{b}}$ containing $E$. So by Corollary \ref{c26}  there is $A\in \binom{U_{q+1}}{6}$ such that $B\subseteq A$ and $\sigma_{6,3}(A)=0$. Thus $A\in \mathcal{T}_{1}\cup \mathcal{T}_{2}$, where $\mathcal{T}_{1}$ and $\mathcal{T}_{2}$ are the same defined as in Corollary \ref{c26}. Let
$$\mathcal{B}_{i}=\Bigg\{B\in \binom{U_{q+1}}{5}: \text{there is} \ A\in \mathcal{T}_{i}\ \text{such that}\ E\subseteq B\subseteq A\Bigg\},$$
where $i=1,2$. Clearly $\mathcal{B}_{1}$ and $\mathcal{B}_{2}$ are disjoint.

For any $A\in \mathcal{T}_{1}$, there exists $u\in A\setminus E$ such that $\sigma_{5,2}(A\setminus\{u\})=0$. From Lemma \ref{lemma23} there is a unique $B_{0}\in \binom{U_{q+1}}{5}$ such that $E\subseteq B_{0}$ and $\sigma_{5,2}(B_{0})=0$. Hence $A\setminus \{u\}=B_0$. Let $B_0=\{u_{1},u_{2},u_{3},u_{4},u_{5}\}$. By Lemma \ref{l3.3}, clearly $B_0\in \mathcal{B}_{1}$ and also $\{u_{1},u_{2},u_{3},u_{i},\alpha\}\in \mathcal{B}_1$ for any $i=4,5$ and $\alpha\in U_{q+1}\setminus B_{0}$. As a consequence, each $A\in \mathcal{T}_{1}$ provides $1+2(q-4)$ distinct blocks of $\mathcal{B}_{1}$ and hence $|\mathcal{B}_{1}|=1+2(q-4)$.

For any $A\in \mathcal{T}_{2}$, obviously $A\setminus \{u\}\in \mathcal{B}_{2}$ for any $u\in A\setminus E$. From Corollary \ref{c26} (2), $|\mathcal{T}_{2}|=\frac{(q-4)(q-10)}{6}$. Applying Corollary \ref{c27} (2) yields that each $A\in \mathcal{T}_{2}$ provides three distinct blocks of $\mathcal{B}_{2}$. Hence $|\mathcal{B}_{2}|=3|\mathcal{T}_{2}|=\frac{(q-4)(q-10)}{2}$.

Now we conclude that $|\mathcal{B}_{1}|+|\mathcal{B}_{2}|=\frac{q^2-10q+26}{2}$ and $(U_{q+1},\mathcal{B}^{\overline{b}}_{\sigma_{5,3},q+1})$ is a $3$-$(q+1,5,\frac{q^2-10q+26}{2})$ design.\qed

\begin{theorem}\label{t36}
For $q=2^{m}$ with even $m \geqslant 4$, $(U_{q+1},\mathcal{B}^{b}_{\sigma_{6,2},q+1})$ forms a $3$-$(q+1,6,2q-8)$ design.
\end{theorem}

\proof Let $B\in \binom{U_{q+1}}{6}$. By Eq. (\ref{Bu}), $B \in \mathcal{B}^{b}_{\sigma_{6,2},q+1}$ if and only if $\sigma_{6,2}(B-a)=0$ for some $a\in B$. From Eq. (\ref{ba}) one has
$$\sigma_{6,2}(B-a) =\sigma_{6,2}(B)-5a\sigma_{6,1}(B)+15a^{2}=\sigma_{6,2}(B)+a\sigma_{6,1}(B)+a^{2}.$$
It is clear by noting $a\in B$ that
$$\left.
    \begin{array}{l}
    \sigma_{6,2}(B)+a\sigma_{6,1}(B)+a^{2} \\
        =\sigma_{5,2}(B \setminus\{a\})+a\sigma_{5,1}(B \setminus\{a\})+a(\sigma_{5,1}(B \setminus\{a\})+a)+a^{2}\\
       =\sigma_{5,2}(B \setminus\{a\}).  \\
    \end{array}
  \right.
$$
So $\sigma_{6,2}(B-a)=0$ for $a\in B$ is the same as  $\sigma_{5,2}(B \setminus \{a\})=0$. Let $E=\{u_{1},u_{2},u_{3}\} $ be a fixed $3$-subset of $U_{q+1}$ and let $B$ be any block of $\mathcal{B}_{\sigma_{6,2},q+1}^{b}$ containing $E$. Then we have two possibilities for $B$.

Case 1. $\sigma_{5,2}(B\setminus \{a\})=0$ for some $a\in B\setminus E$. Denote the collection of such $B$ by $\mathcal{B}_{1}$. From Lemma \ref{lemma23}, $(U_{q+1},\mathcal{B}_{\sigma_{5,2},q+1})$ is an $\Sx(3,5,q+1)$. Then $|\{A\in \mathcal{B}_{\sigma_{5,2},q+1}: E\subseteq A\}|=1$. From Corollary \ref{c27} (1), $|A_{1}\cap A_{2}|\leqslant 2$ for any $A_{1}$, $A_{2}\in \mathcal{B}_{\sigma_{5,2},q+1}$ with $A_{1}\neq A_{2}$. Then $\mathcal{B}_{1}$ is simple and $|\mathcal{B}_{1}|=| \{A\cup \{u\}: A\in \mathcal{B}_{\sigma_{5,2},q+1}, E \subseteq A, u\in U_{q+1} \setminus A\}|=q+1-5=q-4$.

Case 2. $\sigma_{5,2}(B\setminus \{a\})=0$ for some $a\in E$. Denote the collection of such $B$ by $\mathcal{B}_{2}$. Also note that $(U_{q+1},\mathcal{B}_{\sigma_{5,2},q+1})$ is an $\Sx(3,5,q+1)$. Then there exists a unique block of $\mathcal{B}_{\sigma_{5,2},q+1}$ containing $E$. Applying Eq. (\ref{1}), there are $\lambda_{2}=\frac{q+1-2}{3}$ blocks of $\mathcal{B}_{\sigma_{5,2},q+1}$ containing any fixed pair of $E$. Excluding the unique block containing $E$ and applying Corollary \ref{c27} (1), we have that $\mathcal{B}_{2}$ is simple and $|\mathcal{B}_{2}|=3(\frac{q-1}{3}-1)=q-4$.

Now we prove $\mathcal{B}_{1}\cap \mathcal{B}_{2}=\emptyset$. Otherwise, there exists $B\in \mathcal{B}_{1}\cap \mathcal{B}_{2}$. Then $\sigma_{5,2}(B\setminus \{u_{i}\})=0$ for some $u_{i}\in B\setminus E$ and $\sigma_{5,2}(B\setminus \{u_{j}\})=0$ for some $u_{j}\in E$. Note that $|(B\setminus \{u_{i}\})\cap (B\setminus \{u_{j}\})|= 4$, contradicting Corollary \ref{c27} (1). Then $\mathcal{B}_{1}$ and $\mathcal{B}_{2}$ are disjoint. So $E$ is contained in $|\mathcal{B}_{1}|+|\mathcal{B}_{2}|=2(q-4)$ blocks of $\mathcal{B}^{b}_{\sigma_{6,2},q+1}$ and $(U_{q+1},\mathcal{B}^{b}_{\sigma_{6,2},q+1})$ is a $3$-$(q+1,6,2q-8)$ design. \qed

\begin{lemma}\label{73u} For $q=2^{m}$, one has
$$\mathcal{B}^{u}_{\sigma_{7,3},q+1}=\bigg\{B\in \binom{U_{q+1}}{7}:\sigma_{6,3}(B\setminus \{a\})=0 \ \text{for some}\ a\in B\bigg\}.$$
\end{lemma}

\proof Let $B=\{ u_{1},u_{2},\ldots,u_{7}\}$ be any block of $\mathcal{B}^{u}_{\sigma_{7,3},q+1}$. By Eq. (\ref{Bu}),  $\sigma_{7,3}(B-a)=0$ for some $a\in U_{q+1}$. From Eq. (\ref{ba}) one obtains
\begin{eqnarray*}
 &&\sigma_{7,3}(B-a)=\sigma_{7,3}(B)-5a\sigma_{7,2}(B)+15a^{2}\sigma_{7,1}(B)-35a^{3}\\
&& =\sigma_{7,3}(B)+a\sigma_{7,2}(B)+a^{2}\sigma_{7,1}(B)+a^{3}=0.
\end{eqnarray*}
Next we prove $a \in B$.  Note that
$$\left.
    \begin{array}{l}
       a^{3}\sigma_{7,4}+a^{2}\sigma_{7,5}+a\sigma_{7,6}+\sigma_{7,7}
    \\=a^{3}\sigma_{7,7}\sigma_{7,3}^{q}+a^{2}\sigma_{7,7}\sigma_{7,2}^{q}+a\sigma_{7,7}\sigma_{7,1}^{q}+\sigma_{7,7}  \\
       =a^{3}\sigma_{7,7}(\sigma_{7,3}^{q}+a^{-1}\sigma_{7,2}^{q}+a^{-2}\sigma_{7,1}^{q}+a^{-3})  \\
       =a^{3}\sigma_{7,7}(\sigma_{7,3}^{q}+a^{q}\sigma_{7,2}^{q}+a^{2q}\sigma_{7,1}^{q}+a^{3q})  \\
       =a^{3}\sigma_{7,7}(\sigma_{7,3}+a\sigma_{7,2}+a^{2}\sigma_{7,1}+a^{3})^{q} \\
       =0.\\
    \end{array}
  \right.
$$
Multiplying both sides of $\sigma_{7,3}+a\sigma_{7,2}+a^{2}\sigma_{7,1}+a^{3}=0$ by $a^4$ and then adding $a^{3}\sigma_{7,4}+a^{2}\sigma_{7,5}+a\sigma_{7,6}+\sigma_{7,7}$ yields
$$\sigma_{7,7}+a\sigma_{7,6}+a^{2}\sigma_{7,5}+a^{3}\sigma_{7,4}+a^{4}\sigma_{7,3}+a^{5}\sigma_{7,2}+a^{6}\sigma_{7,1}+a^{7}=0.$$
So
$$(a+u_{1})(a+u_{2})(a+u_{3})(a+u_{4})(a+u_{5})(a+u_{6})(a+u_{7})=0.$$
Thus $a=u_{i}$ for some $u_{i}\in B$ and
$$\left.
    \begin{array}{l}
      \sigma_{7,3}(B-u_{i}) \\ = \sigma_{7,3}+u_{i}\sigma_{7,2}+u_{i}^{2}\sigma_{7,1}+u_{i}^{3}\\
       =(\sigma_{6,3}(B\setminus \{u_{i}\})+u_{i}\sigma_{6,2}(B\setminus \{u_{i}\}) )+u_{i}(\sigma_{6,2}(B\setminus \{u_{i}\})\\ \quad  +u_{i}\sigma_{6,1}(B\setminus \{u_{i}\}))+u_{i}^{2}(\sigma_{6,1}(B\setminus \{u_{i}\})+u_{i})+u_{i}^{3} \\
       =\sigma_{6,3}(B\setminus \{u_{i}\})+2(u_{i}\sigma_{6,2}(B\setminus \{u_{i}\})+u_{i}\sigma_{6,1}(B\setminus \{u_{i}\})+u_{i}^{3})\\
       =\sigma_{6,3}(B\setminus \{u_{i}\}).  \\
    \end{array}
  \right.
$$
Hence the conclusion follows.\qed

\begin{theorem}\label{t3.7}
For $q=2^{m}$ with odd $m \geqslant 5$,  $(U_{q+1},\mathcal{B}^{u}_{\sigma_{7,3},q+1})$ forms a $4$-$(q+1,7,\frac{7(q-8)(q-5)}{6})$ design.
\end{theorem}

\proof By Lemma \ref{73u}, $\mathcal{B}^{u}_{\sigma_{7,3},q+1}=\{B\in \binom{U_{q+1}}{7}:\sigma_{6,3}(B\setminus \{a\})=0 \ \text{for some}\ a\in B\}.$ Let $E=\{u_{1},u_{2},u_{3},u_{4}\}$ be a fixed 4-subset of $U_{q+1}$ and let $B$ be any block of $\mathcal{B}_{\sigma_{7,3},q+1}^{u}$ containing $E$. Then we have two possibilities for $B$.

Case 1. $\sigma_{6,3}(B \setminus \{a\})=0$ for some $a\in B\setminus E$. Denote the  collection of such $B$ by $\mathcal{B}_{1}$. From Lemma \ref{lemma 24}, $(U_{q+1},\mathcal{B}_{\sigma_{6,3},q+1})$ is a $4$-$(q+1,6,\frac{q-8}{2})$ design. Then $|\{A\in \mathcal{B}_{\sigma_{6,3},q+1}: E\subseteq A\}|=\frac{q-8}{2}$. From Corollary \ref{c27} (3), $|A_{1}\cap A_{2}|\leqslant 4$ for any $A_{1}$, $A_{2}\in \mathcal{B}_{\sigma_{6,3},q+1}$ with $A_{1}\neq A_{2}$. Then $\mathcal{B}_{1}$ is simple and
$|\mathcal{B}_{1}|=| \{A\cup \{u\}: A\in \mathcal{B}_{\sigma_{6,3},q+1}, E \subseteq A, u\in U_{q+1} \setminus A\}|=\frac{q-8}{2}\cdot (q+1-6)=\frac{(q-5)(q-8)}{2}$.

Case 2. $\sigma_{6,3}(B \setminus \{a\})=0$ for some $a\in E$. Denote the  collection of such $B$ by $\mathcal{B}_{2}$. By Lemma \ref{lemma 24}, there are $\frac{q-8}{2}$ blocks of $\mathcal{B}_{\sigma_{6,3},q+1}$ containing $E$. Applying Eq. (\ref{1}), there are $\lambda_{3}=\frac{q-8}{2}\cdot \frac{q+1-3}{3}$ blocks of $\mathcal{B}_{\sigma_{6,3},q+1}$ containing any fixed $3$-subset of $E$. Excluding $\frac{q-8}{2}$ blocks containing $E$ and applying Corollary \ref{c27} (3) yields that $\mathcal{B}_{2}$ is simple and $|\mathcal{B}_{2}|=4(\lambda_{3}-\frac{q-8}{2})=\frac{2(q-8)(q-5)}{3}$.

Now we prove $\mathcal{B}_{1}\cap \mathcal{B}_{2}=\emptyset$. Otherwise, there exists $B\in \mathcal{B}_{1}\cap \mathcal{B}_{2}$. Then $\sigma_{6,3}(B \setminus \{u_{i}\})=0$ for some $u_{i}\in B\setminus E$ and $\sigma_{6,3}(B \setminus \{u_{j}\})=0$ for some $u_{j}\in E$. Note that $|(B\setminus \{u_{i}\})\cap (B\setminus \{u_{j}\})|= 5$, contradicting Corollary \ref{c27} (3). Hence $\mathcal{B}_{1}$ and $\mathcal{B}_{2}$ are disjoint. So $E$ is contained in $|\mathcal{B}_{1}|+|\mathcal{B}_{2}|=\frac{7(q-8)(q-5)}{6}$ blocks of $\mathcal{B}^{u}_{\sigma_{7,3},q+1}$ and thus $(U_{q+1},\mathcal{B}^{u}_{\sigma_{7,3},q+1})$ is a $4$-$(q+1,7,\frac{7(q-8)(q-5)}{6})$ design. \qed

\begin{theorem}\label{t3.8}
For $q=2^{m}$ with even $m \geqslant 4$,  $(U_{q+1},\mathcal{B}^{u}_{\sigma_{7,3},q+1})$ is a $3$-$(q+1,7,\frac{7(q-4)(q-5)(q-10)}{24})$ design.
\end{theorem}

\proof Let $B\in \binom{U_{q+1}}{7}$. By Lemma \ref{73u}, $B \in \mathcal{B}^{u}_{\sigma_{7,3},q+1}$ if and only if $\sigma_{6,3}(B\setminus \{a\})=0$ for some $a\in B$. Let $E=\{u_{1},u_{2},u_{3}\}$ be a fixed 3-subset of $U_{q+1}$. Let $B=\{u_{1},u_{2},\ldots,u_{7}\}$ be any block of $\mathcal{B}_{\sigma_{7,3},q+1}^{u}$ containing $E$. We treat four possibilities of $B$ according to Corollary \ref{c26}. Clearly $B\in \bigcup_{i=1}^{4}\mathcal{B}_{i}$, where
$$\left.
    \begin{array}{l}
    \mathcal{B}_{1}=\left \{
                    \begin{array}{cl}
                     B\in \binom{U_{q+1}}{7}:  &E \subseteq B, \sigma_{6,3}(B\setminus \{u_{i}\})=0\ \text{for some}\  u_{i}\in B\setminus E,\\
                       & \sigma_{5,2}(B \setminus\{u_{i},u_{j}\})=0\ \text{for some} \ u_{j}\in B \setminus (E\cup\{u_{i}\})\\
                    \end{array}
                  \right\},   \\
                  \\
       \mathcal{B}_{2}=\left \{
                    \begin{array}{cl}
                     B\in \binom{U_{q+1}}{7}:  &E \subseteq B, \sigma_{6,3}(B\setminus \{u_{k}\})=0\ \text{for some}\  u_{k}\in E,\\
                       & \sigma_{5,2}(B \setminus\{u_{k},u_{j}\})=0\ \text{for some} \ u_{j}\in B \setminus E\\
                    \end{array}
                  \right\},\\
                  \\
     \mathcal{B}_{3}=\left \{
                    \begin{array}{cl}
                     B\in \binom{U_{q+1}}{7}:  &E \subseteq B, \sigma_{6,3}(B\setminus \{u_{i}\})=0\ \text{for some}\  u_{i}\in  B \setminus E,\\
                       & \sigma_{5,2}(B \setminus\{u_{i},u_{j}\})\ne 0\ \text{for any} \ u_{j}\in B \setminus (E\cup\{u_{i}\})\\
                    \end{array}
                  \right\},\\
                  \\
     \mathcal{B}_{4}=\left \{
                    \begin{array}{cl}
                     B\in \binom{U_{q+1}}{7}:  &E \subseteq B, \sigma_{6,3}(B\setminus \{u_{k}\})=0\ \text{for some}\  u_{k}\in E,\\
                       & \sigma_{5,2}(B \setminus\{u_{k},u_{j}\})\ne 0\ \text{for any} \ u_{j}\in B \setminus E\\
                    \end{array}
                  \right\}.
    \end{array}
  \right.
$$
Next we prove that $\mathcal{B}_{2}\subseteq \mathcal{B}_{3}$ and that $\mathcal{B}_{1}$, $\mathcal{B}_{3}$ and $\mathcal{B}_{4}$ are mutually disjoint.
\begin{enumerate}
\item[(i)] For any $B\in \mathcal{B}_{2}$, w.l.o.g. we assume $\sigma_{6,3}(B\setminus \{u_{1}\})=0$ and $\sigma_{5,2}(B\setminus\{u_{1},u_{7}\})=0$. Then we have $\sigma_{6,3}(B\setminus\{u_{7}\})=0$ by Lemma \ref{l3.3}.
Note that $|(B\setminus\{u_{1},u_{7}\})\cap(B\setminus\{u_{7},u_{j}\})| =4$ for any $u_{j}\in B\setminus E$. Then from Corollary \ref{c27} (1), $\sigma_{5,2}(B\setminus\{u_{7},u_{j}\})\ne 0$ as $\sigma_{5,2}(B\setminus\{u_{1},u_{7}\})=0$. This shows that $B\in \mathcal{B}_{3}$ by taking $u_{i}=u_{7}$. As a result, $\mathcal{B}_{2} \subseteq \mathcal{B}_{3}$.
\item[(ii)] For any $B\in \mathcal{B}_{1}\cap \mathcal{B}_{3}$, w.l.o.g. we assume $\sigma_{6,3}(B\setminus\{u_{7}\})=0$, $\sigma_{5,2}(B\setminus\{u_{6},u_{7}\})=0$, $\sigma_{6,3}(B\setminus \{u_{i}\})=0$ for some $u_{i}\in B\setminus E$ and $\sigma_{5,2}(B\setminus \{u_{i},u_{j}\})\ne 0$ for any $u_{j}\in B\setminus (E\cup \{u_{i}\})$. It is obvious that $u_{i}\not \in \{u_{6},u_{7}\}$ and $|(B\setminus\{u_{i}\})\cap(B\setminus\{u_{7}\})|=5$. So we have $\sigma_{5,2}(B\setminus \{u_{i},u_{7}\})=0$ from Corollary \ref{c27} (2), which contradicts $B\in \mathcal{B}_{3}$. Hence $\mathcal{B}_{1}\cap \mathcal{B}_{3}=\emptyset$.

\end{enumerate}
Similarly, we can prove that $\mathcal{B}_{1} \cap \mathcal{B}_{4}= \emptyset$ and $\mathcal{B}_{3} \cap \mathcal{B}_{4}= \emptyset$. To prove the finial conclusion we only need to show that $|\mathcal{B}_{1}|+|\mathcal{B}_{3}|+|\mathcal{B}_{4}|=\frac{7(q-4)(q-5)(q-10)}{24}$. Next we calculate the cardinalities of $\mathcal{B}_{1}$, $\mathcal{B}_{3}$ and $\mathcal{B}_{4}$.

By Lemma \ref{lemma23}, there is a unique block $A\in \binom{U_{q+1}}{5}$ such that $E\subseteq A$ and $\sigma_{5,2}(A)=0$. From Lemma \ref{l3.3}, $\sigma_{6,3}(A\cup\{u\})=0$ for any $u\in U_{q+1}\setminus A$. This implies $B\in \mathcal{B}_{1}$ if and only if $B=A\cup\{u_{6},u_{7}\}$ where $\{u_{6},u_{7}\}\in \binom{U_{q+1}\setminus A}{2}$. It then follows that  \begin{equation}\label{B1}|\mathcal{B}_{1}|=\frac{(q-4)(q-5)}{2}.\end{equation}

It is clear that $B\in\mathcal{B}_{3}$ if and only if $B=B'\cup \{u\}$ for some $B'\in \mathcal{B}'$ and $u\in U_{q+1}\setminus B'$, where $$\mathcal{B}'=\bigg\{B'\in \binom{U_{q+1}}{6}: E \subseteq B', \sigma_{6,3}(B')=0, \sigma_{5,2}(B'\setminus \{u'\})\ne 0 \ \text{for any}\ u'\in B'\setminus E\bigg\}.$$
From Corollary \ref{c26} (2), we have $|\mathcal{B}'|=\frac{(q-4)(q-10)}{6}$. And from Corollary \ref{c27} (2), $|B'_{1}\cap B'_{2}|\leqslant 4$ for any $B'_{1}$, $B'_{2}\in \mathcal{B}_{\sigma_{6,3},q+1}$ with $B'_{1}\neq B'_{2}$. It then follows that \begin{equation}\label{B3}|\mathcal{B}_{3}|=\frac{(q-4)(q-10)}{6} \cdot(q+1-6)=\frac{(q-4)(q-10)(q-5)}{6}.\end{equation}

In order to calculate the size of $\mathcal{B}_{4}$, we define $\mathcal{T}_{k,1}$ and $\mathcal{T}_{k,2}$ for any fixed $u_{k}\in E$ by
$$\mathcal{T}_{k,1}=\bigg\{B\in \binom{U_{q+1}}{7}: E\subseteq B, \sigma_{6,3}(B \setminus \{u_{k}\})=0\bigg\},$$
and
$$\mathcal{T}_{k,2}=\bigg\{B\in \binom{U_{q+1}}{7}: E\subseteq B, \sigma_{5,2}(B\setminus \{u_{k},u_{j}\})=0\ \text{for some} \ u_{j}\in B \setminus E\bigg\}.$$
Then $\mathcal{T}_{k,2}\subseteq \mathcal{T}_{k,1}$ by Lemma \ref{l3.3}. Let $\mathcal{T}_{k}=\mathcal{T}_{k,1}\setminus \mathcal{T}_{k,2}$. Clearly, $\mathcal{B}_{4}=\mathcal{T}_{1}\cup\mathcal{T}_{2}\cup\mathcal{T}_{3}$.

It is obvious that $B\in \mathcal{T}_{k,1}$ if and only if $B=B'\cup\{u_{k}\}$ for some $B'\in \mathcal{B}_{\sigma_{6,3},q+1}$ such that $E \cap B'=E\setminus \{u_{k}\}$. From Lemma \ref{l25} and Eq. (\ref{1}), $E$ is contained in $\frac{(q-4)^2}{6}$ blocks of $\mathcal{B}_{\sigma_{6,3},q+1}$ and $E\setminus \{u_{k}\}$ is contained in $\lambda_{2}=\frac{(q-4)^{2}}{6}\cdot\frac{q-1}{4}=\frac{(q-4)^2(q-1)}{24}$ blocks of $\mathcal{B}_{\sigma_{6,3},q+1}$. It follows that $|\mathcal{T}_{k,1}|=\lambda_{2}-\frac{(q-4)^{2}}{6}=\frac{(q-4)^{2}(q-5)}{24}$ as $u_{k}\notin B'$.

Clearly, $B\in \mathcal{T}_{k,2}$ if and only if $B=B'\cup\{u_{k},u_{j}\}$ for some $B'\in \mathcal{B}_{\sigma_{5,2},q+1}$ with $E \cap B'=E\setminus \{u_{k}\}$ and $u_{j}\in U_{q+1}\setminus (B'\cup\{u_{k}\})$. From Lemma \ref{lemma23} and (\ref{1}), we have $\frac{q+1-2}{3}-1=\frac{q-4}{3}$ choices of $B'$ and $|\mathcal{T}_{k,2}|=\frac{(q-4)(q-5)}{3}$ by noting Corollary \ref{c27} (1). Hence, \begin{equation}\label{TK}|\mathcal{T}_{k}|=|\mathcal{T}_{k,1}|-|\mathcal{T}_{k,2}|=\frac{(q-4)(q-5)(q-12)}{24}.\end{equation}

Let $\{u_{i},u_{j}\}\subseteq E$ be fixed and define
$$\mathcal{A}_{ij}=\Big\{B\in \binom{U_{q+1}}{7}: E \subseteq B, \sigma_{5,2}(B\setminus\{u_{i},u_{j}\})=0 \Big\}.$$
Now we prove $\mathcal{T}_{i}\cap \mathcal{T}_{j}=\mathcal{A}_{ij}$. For any $B\in \mathcal{A}_{ij}$, we have $\sigma_{5,2}(B\setminus \{u_{i},u_{j}\})=0$. From Lemma \ref{l3.3}, $\sigma_{6,3}(B\setminus \{u_{i}\})=0$ and $\sigma_{6,3}(B\setminus\{u_{j}\})=0$. Since $|(B\setminus \{u_{i},u_{j}\})\cap(B\setminus \{u_{i},u_{k}\})|=4$ for any $u_{k}\in U_{q+1}\setminus E$, we have $\sigma_{5,2}(B\setminus \{u_{i},u_{k}\})\ne 0$ by Corollary \ref{c27} (1) as $\sigma_{5,2}(B\setminus \{u_{i},u_{j}\})=0$. Thus $B\in \mathcal{T}_{i}$. Similarly, we also have $B\in \mathcal{T}_{j}$ and thus $\mathcal{A}_{ij}\subseteq \mathcal{T}_{i} \cap\mathcal{T}_{j}$. On the other hand, for any $B\in \mathcal{T}_{i}\cap \mathcal{T}_{j}$, we have $\sigma_{6,3}(B\setminus \{u_{i}\})=0$, $\sigma_{6,3}(B\setminus \{u_{j}\})=0$ and $|(B\setminus \{u_{i}\})\cap(B\setminus \{u_{j}\})|=5$. So $\sigma_{5,2}(B\setminus\{u_{i},u_{j}\})=0$ from Corollary \ref{c27} (2). Then $B\in \mathcal{A}_{ij}$. Clearly $\mathcal{T}_{1}\cap \mathcal{T}_{2}\cap\mathcal{T}_{3}=\emptyset$.

Now we calculate the size of $\mathcal{A}_{ij}$ $(1\leqslant i<j\leqslant 3)$.  From Lemma \ref{lemma23}, there is a unique block of $\mathcal{B}_{\sigma_{5,2},q+1}$ containing $E$. By Lemma \ref{lemma23} and Eq. (\ref{1}), there are $\lambda_{1}=\frac{\binom{q+1-1}{3-1}}{\binom{5-1}{3-1}}=\frac{q(q-1)}{12}$ blocks of $\mathcal{B}_{\sigma_{5,2},q+1}$ containing any fixed point of $ E$ and $\lambda_{2}=\frac{q-1}{3}$ blocks of $\mathcal{B}_{\sigma_{5,2},q+1}$ containing any fixed pair of $E$. Applying Corollary \ref{c27} (1) and the principle of inclusion-exclusion,  \begin{equation}\label{Aij}|\mathcal{A}_{ij}|=\lambda_{1}-2\lambda_{2}+1=\frac{(q-4)(q-5)}{12}.\end{equation}

Since $\mathcal{B}_4=\mathcal{T}_{1}\cup \mathcal{T}_{2}\cup \mathcal{T}_{3}$, $\mathcal{T}_{i}\cap \mathcal{T}_{j}=\mathcal{A}_{ij}$, and $\mathcal{T}_{1}\cap \mathcal{T}_{2}\cap \mathcal{T}_{3}=\emptyset$, using (\ref{TK}) and (\ref{Aij}) gives
\begin{equation}\label{B4}|\mathcal{B}_{4}|=\sum\limits^{3}_{k=1}|\mathcal{T}_{k}|-\sum\limits_{1\leqslant i< j \leqslant 3} |\mathcal{A}_{ij}|=\frac{(q-4)(q-5)(q-14)}{8}.\end{equation}
Combining (\ref{B1}),   (\ref{B3}) and  (\ref{B4}) yields that $E$ is contained in $|\mathcal{B}_{1}|+|\mathcal{B}_{3}|+|\mathcal{B}_{4}|=\frac{7(q-4)(q-5)(q-10)}{24}$ blocks of $\mathcal{B}^{u}_{\sigma_{7,3},q+1}$. Consequently, $(U_{q+1},\mathcal{B}^{u}_{\sigma_{7,3},q+1})$ is a $3$-$(q+1,7,\frac{7(q-4)(q-5)(q-10)}{24})$ design.\qed

From Theorems \ref{t3.7} and \ref{t3.8}, one gets
\begin{equation}\label{e152}
|\mathcal{B}_{\sigma_{7,3},q+1}^{u}|=\left\{
                                              \begin{array}{ll}
  \frac{(q-4)(q-5)(q-10)}{120}\binom{q+1}{3}, & \text{if}\  q=2^{2s}, \\
 \frac{(q-5)(q-8)}{30}\binom{q+1}{4},   & \text{if} \ q=2^{2s+1}. \\
                                              \end{array}
                                            \right.
\end{equation}
In general, it is difficult to determine $|\mathcal{B}_{\sigma_{k,l},q+1}^{u}|$. It would be interesting to settle the following problem.

\begin{problem}
Determine the cardinality of  $\mathcal{B}_{\sigma_{k,l},q+1}^{u}$ for $(k,l)\ne (4,2),(5,2), (5,3), (6,3),(7,3)$.
\end{problem}

%
%

\begin{theorem}\label{t3.2222}
Let $q=2^{m}$ and $m \geqslant 4$ be even. Then  $(U_{q+1},\mathcal{B}^{0}_{\sigma_{7,3},q+1})$ forms a $3$-$(q+1,7,\frac{7(q-5)(q-4)}{4})$ design, where the block set $\mathcal{B}^{0}_{\sigma_{7,3},q+1}$ is given by
$$\mathcal{B}^{0}_{\sigma_{7,3},q+1}=\Bigg\{B\in\binom{U_{q+1}}{7}:
                      \sigma_{5,2}(B\setminus\{u_{i},u_{j}\})=0\ \text{for some}\ u_{i}\ne u_{j} \in B\Bigg\}.$$
\end{theorem}

\proof Let $E=\{u_{1},u_{2},u_{3}\}$ be a fixed $3$-subset of $U_{q+1}$ and let $B$  be any block of $\mathcal{B}^{0}_{\sigma_{7,3},q+1}$ containing $E$. Then we have three possibilities for $B$.

Case 1. $\sigma_{5,2}(B \setminus\{u_{i},u_{j}\})=0$ for some $ u_{i},u_{j}\in  B\setminus E$ ($u_{i}\ne u_{j}$). Denote the  collection of such $B$ by $\mathcal{B}_{1}$. Clearly $B\in \mathcal{B}_{1}$ if and only if $B=A\cup \{u_{i},u_{j}\}$ for $A\in \mathcal{B}_{\sigma_{5,2},q+1}$ with $E\subseteq A$ and $\{u_{i},u_{j}\}\subseteq \binom{U_{q+1}\setminus A}{2}$. By Lemma \ref{lemma23} and Corollary \ref{c27} (1), we have that $\mathcal{B}_{1}$ is simple and $|\mathcal{B}_{1}|=\frac{(q-4)(q-5)}{2}$.

Case 2. $\sigma_{5,2}(B\setminus\{u_{k},u_{j}\})=0$ for some $u_{k}\in E$ and $u_{j}\in B\setminus E$. Denote the  collection of such $B$ by $\mathcal{B}_{2}$. Clearly $B\in \mathcal{B}_{2}$ if and only if $B=A\cup \{u_{k},u_{j}\}$ for $ A\in \mathcal{B}_{\sigma_{5,2},q+1}$ with $E \cap A=E\setminus\{u_{k}\}$ and $u_{j}\in U_{q+1}\setminus (A\cup\{u_{k}\})$. By Lemma \ref{lemma23} and Eq. (\ref{1}), we have $\lambda_{2}-1=\frac{q+1-2}{3}-1=\frac{q-4}{3}$ choices of $A$ as $u_{k}\notin A$. Then $\mathcal{B}_{2}$ is simple and $|\mathcal{B}_{2}|=3\cdot \frac{q-4}{3}\cdot(q+1-6)=(q-4)(q-5)$  from Corollary \ref{c27} (1).

Case 3. $\sigma_{5,2}(B \setminus\{u_{k},u_{l}\})=0$ for some $u_{k},u_{l}\in E$ ($u_{k}\ne u_{l}$). Denote the  collection of such $B$ by $\mathcal{B}_{3}$. Clearly $B\in \mathcal{B}_{3}$ if and only if $B=A\cup \{u_{k},u_{l}\}$ for $A\in \mathcal{B}_{\sigma_{5,2},q+1}$ with $E\cap A=E\setminus \{u_{k},u_{l}\}$. By Lemma \ref{lemma23} and Eq. (\ref{1}), we have $\lambda_{1}-2\lambda_{2}+1=\frac{\binom{q+1-1}{3-1}}{\binom{5-1}{3-1}}-2\cdot\frac{q-1}{3}+1=\frac{(q-4)(q-5)}{12}$ choices of $A$ as $\{u_{k},u_{l}\}\not\subseteq A$. Then $\mathcal{B}_{3}$ is simple and $|\mathcal{B}_{3}|=3\cdot\frac{(q-4)(q-5)}{12}=\frac{(q-4)(q-5)}{4}$ from Corollary \ref{c27} (1).

Now we prove $\mathcal{B}_{1}\cap \mathcal{B}_{2}=\emptyset$. Otherwise, there is $B\in \mathcal{B}_{1}\cap \mathcal{B}_{2}$. Then $\sigma_{5,2}(B\setminus \{u_{i},u_{j}\})=0$ and $\sigma_{5,2}(B\setminus \{u_{k},u_{l}\})=0$ for $u_{i}, u_{j}, u_{l}\in B\setminus E$ and $u_{k}\in E$. Note that $|(B\setminus \{u_{i},u_{j}\})\cap (B\setminus \{u_{k},u_{l}\})|=3$ or $4$, which contradicts Corollary \ref{c27} (1). Then $\mathcal{B}_{1}$ and $\mathcal{B}_{2}$ are disjoint. Similarly, $\mathcal{B}_{1}$, $\mathcal{B}_{2}$ and $\mathcal{B}_{3}$ are pairwise disjoint. So $E$ is contained in $|\mathcal{B}_{1}|+|\mathcal{B}_{2}|+|\mathcal{B}_{3}|=\frac{7(q-4)(q-5)}{4}$ blocks of $\mathcal{B}^{0}_{\sigma_{7,3},q+1}$ and thus $(U_{q+1},\mathcal{B}^{0}_{\sigma_{7,3},q+1})$ is a $3$-$(q+1,7,\frac{7(q-4)(q-5)}{4})$ design. \qed

\section{BCH codes supporting $t$-designs}
In this section, we consider the codewords of weight $7$ in the narrow-sense BCH codes $\mathcal{C}_{(q,q+1,4,1)}$ over GF$(q)$ where $q=2^m$. We will prove that ${\cal B}_7(\mathcal{C}_{(q,q+1,4,1)})$ supports a $4$-design when $m \geqslant 5$ is odd and it supports a $3$-design when $m \geqslant 4$ is even.

\begin{lemma}{\rm\cite[Theorems 34, 35]{Tang2020An}}\label{l4.1}
Let $q=2^{m}$ and $m\geqslant 4$. Then $\mathcal{C}_{(q,q+1,4,1)}$ has parameters $[q+1,q-5,d]$ with $d=6$ if $m$ is odd and $d=5$ if $m$ is even. The dual $\mathcal{C}_{(q,q+1,4,1)}^{\bot}$ has parameters $[q+1,6,q-5]$. In particular, $\mathcal{C}_{(q,q+1,4,1)}$ is an NMDS code if $m$ is odd and A$^{2}$MDS code if $m$ is even.
\end{lemma}

We display here the connections between $\mathcal{B}_{k}(\mathcal{C}_{(q,q+1,4,1)})$ and $\mathcal{B}_{\sigma_{k,l},q+1}$ for $k\in \{5,6,7\}$.

\begin{lemma}\label{l4,3}
Let $q=2^{m}$ with $m\geqslant 4$ and denote $\mathcal{C}:=\mathcal{C}_{(q,q+1,4,1)}$.  If we index the coordinates of the codewords in $\mathcal{C}$ with the elements in $U_{q+1}$, then we may have the following.
\begin{enumerate}
  \item[$(i)$] For odd $m$,  $\mathcal{B}_{6}(\mathcal{C})=\mathcal{B}_{\sigma_{6,3},q+1}$.
  \item[$(ii)$] For even $m$,  $\mathcal{B}_{5}(\mathcal{C})=\mathcal{B}_{\sigma_{5,2},q+1}$ and $\mathcal{B}_{6}(\mathcal{C})=\mathcal{B}_{\sigma_{6,3},q+1} \setminus \mathcal{B}_{\sigma_{6,3},q+1}^{0}$ $($see $(\ref{0}))$.
 \item[$(iii)$]    $\mathcal{B}_{7}(\mathcal{C})\subseteq \mathcal{\overline{B}}^{u}_{\sigma_{7,3},q+1}$, where $\mathcal{\overline{B}}^{u}_{\sigma_{7,3},q+1}=\binom{U_{q+1}}{7}\setminus \mathcal{B}^{u}_{\sigma_{7,3},q+1}$.
\end{enumerate}
\end{lemma}

\proof The first two assertions follow immediately from \rm{\cite[Theorems 36, 39]{Tang2020An}} and their proofs. Now we prove (iii). We denote by $\wt (\mathbf{c})$ and $\Supp (\mathbf{c})$ the weight and the support of a codeword $\mathbf{c}\in \mathcal{C}$, respectively.

 For any $\mathbf{c}\in \mathcal{C}$ with $\wt(\mathbf{c})=7$, denote $\Supp (\mathbf{c})=B=\{u_{1},u_{2},\ldots,u_{7}\}$. Then $B \in \mathcal{B}_{7}(\mathcal{C})$. We will prove that $B\in \mathcal{\overline{B}}^{u}_{\sigma_{7,3},q+1}$ by contradiction. Suppose on the contrary that $B\in \mathcal{B}^{u}_{\sigma_{7,3},q+1}$. Then there is $u_{i}\in B$ such that $\sigma_{6,3}(B\setminus \{u_{i}\})=0$ from Lemma \ref{73u}. W.l.o.g. let $\sigma_{6,3}(B\setminus \{u_{7}\})=0$. Contradictions will be derived by considering the following two cases.

Case 1. Let $m$ be odd.  The assertion (i) shows that there is $\mathbf{c}_{1}\in \mathcal{C}$ such that $\Supp(\mathbf{c}_{1})=B \setminus \{u_{7}\}$ since  $\mathcal{B}_{6}(\mathcal{C})=\mathcal{B}_{\sigma_{6,3},q+1}$. Doing a linear combination of $\mathbf{c}$ and $\mathbf{c}_{1}$ gives a codeword $\mathbf{c}_{2}\in \mathcal{C}$ such that $\Supp(\mathbf{c}_{2})=B\setminus \{u_{1}\}$ and wt$(\mathbf{c}_{2})=6$ as the minimum weight of $\mathcal{C}$ is 6 by Lemma \ref{l4.1}. Note that $\sigma_{6,3}(B\setminus \{u_{1}\})=\sigma_{6,3}(B\setminus \{u_{7}\})=0$ and $|(B\setminus \{u_{1}\})\cap (B\setminus \{u_{7}\})|=5$, contradicting Corollary \ref{c27} (3).

Case 2. Let $m$ be even. We begin with a claim.

\noindent {\bf Claim:} There does not exist $\mathbf{c}_{1}\in \mathcal{C}$ such that $\Supp(\mathbf{c}_{1})=B\setminus \{u_{7}\}$.

If the claim is not true, then let $\mathbf{c}_{1}\in \mathcal{C}$ and $\Supp(\mathbf{c}_{1})=B\setminus \{u_{7}\}$. A linear combination of $\mathbf{c}$ and $\mathbf{c}_{1}$ gives a codeword $\mathbf{c}_{2}\in \mathcal{C}$ such that $u_{7}\in \Supp(\mathbf{c}_{2})\subseteq B\setminus \{u_{1}\}$ and wt$(\mathbf{c}_{2})=5,6$ as the minimum weight of $\mathcal{C}$ is 5 by Lemma \ref{l4.1}. If wt($\mathbf{c}_{2}$)=6, then  $\Supp(\mathbf{c}_{2})=B\setminus \{u_{1}\}$ (similarly to Case 1) and $\sigma_{6,3}(B\setminus \{u_{1}\})=\sigma_{6,3}(B\setminus \{u_{7}\})=0$ and thus $\sigma_{5,2}(B\setminus \{u_{1},u_{7}\})=0$ by Corollary \ref{c27} (2), contradicting $\mathbf{c}_{2}\in \mathcal{B}_{6}(\mathcal{C})=\mathcal{B}_{\sigma_{6,3},q+1} \setminus \mathcal{B}_{\sigma_{6,3},q+1}^{0}$ by  (ii). So we must have wt($\mathbf{c}_{2})=5$. Suppose $\Supp(\mathbf{c}_{2})=B\setminus\{u_{1},u_{2}\}$. So $\sigma_{5,2}(B\setminus\{u_{1},u_{2}\})=0$ by  (ii). It is immediate that $\sigma_{6,3}(B \setminus \{u_{1}\})=\sigma_{6,3}(B \setminus \{u_{2}\})=0$ by Lemma \ref{l3.3}. Since we also have $\sigma_{6,3}(B \setminus \{u_{7}\})=0$, combining together and also noting $|(B\setminus \{u_i\})\cap (B\setminus \{u_j\})|=5$ $(i\ne j\in\{1,2,7\})$ yields from Corollary \ref{c27} (2) that $\sigma_{5,2}(B\setminus \{u_{1},u_{7}\})=\sigma_{5,2}(B\setminus \{u_{2},u_{7}\})=0$, but this contradicts Corollary \ref{c27} (1). This completes the proof of the claim.

Applying the claim we know that $B\setminus \{u_{7}\}$ dose not support any codeword of $\mathcal{C}$. Since $\sigma_{6,3}(B\setminus\{u_{7}\})=0$, there exists a $5$-subset $A\subseteq B\setminus\{u_{7}\}$ such that $\sigma_{5,2}(A)=0$ by  (ii), say, $A=\{u_{1},u_{2},u_{3},u_{4},u_{5}\}$. Again from  (ii), there is $\mathbf{c}_{1}\in \mathcal{C}$ such that $\Supp(\mathbf{c}_{1})=A$. A linear combination of $\mathbf{c}$ and $\mathbf{c}_{1}$ gives a codeword $\mathbf{c}_{2}\in \mathcal{C}$ such that $\{u_{6},u_{7}\}\subseteq \Supp(\mathbf{c}_{2})\subseteq B \setminus\{u_{1}\}$ and wt($\mathbf{c}_{2})=5,6$ as the minimum weight of $\mathcal{C}$ is 5 by Lemma \ref{l4.1}. We have wt($\mathbf{c}_{2})=6$ because if wt($\mathbf{c}_{2})=5$ then $\sigma_{5,2}(\Supp(\mathbf{c}_{1}))=\sigma_{5,2}(\Supp(\mathbf{c}_{2}))=0$ by  (ii) and $|\Supp(\mathbf{c}_{1})\cap \Supp(\mathbf{c}_{2})|=3$, contradicting Corollary \ref{c27} (1). Thus $|\Supp(\mathbf{c}_{2})|=6$ and Supp$(\mathbf{c}_{2})=B\setminus \{u_{1}\}$. By assertion (ii), $\sigma_{6,3}(B\setminus\{u_{1}\})=0$. So we have $\sigma_{5,2}(B\setminus\{u_{1},u_{7}\})=0$ by Corollary \ref{c27} (2) as $\sigma_{6,3}(B\setminus\{u_{7}\})=0$. However, this contradicts that $\mathbf{c}_{2}\in \mathcal{B}_{6}(\mathcal{C})=\mathcal{B}_{\sigma_{6,3},q+1} \setminus \mathcal{B}_{\sigma_{6,3},q+1}^{0}$ by  (ii). \qed

\begin{lemma}\label{4.4}
Let $q=2^{m}$ with $m\geqslant 4$ and $\mathcal{C}$ be the narrow-sense BCH code $\mathcal{C}_{(q,q+1,4,1)}$ over $\GF(q)$ with the minimum weight $d$. Then $| \mathcal{B}_{i}(\mathcal{C})|=\frac{A_{i}}{q-1}$ for $d \leqslant i \leqslant 7$, where $A_i$ denotes the number of codewords with weight $i$ in ${\cal C}$.
\end{lemma}

\proof From Lemma \ref{l4.1}, we have $d=5$ if $m$ is even and $d=6$ if $m$ is odd. We only need to prove that if  $\mathbf{c}\in \mathcal{C}$ and $\mathbf{c}_{1} \in \mathcal{C}$ are two codewords of weight $i$ ($d \leqslant i \leqslant 7$) with Supp$(\mathbf{c})$=Supp$(\mathbf{c}_{1})$, then $\mathbf{c}=\alpha \mathbf{c}_{1}$ for some nonzero $\alpha \in $GF$(q)$. The proof will proceed based on the notation and three assertions of Lemma \ref{l4,3}.

 Assume that $\mathbf{c}$ and $\mathbf{c}_{1}$ are two different codewords of $\mathcal{C}$ such that $\wt(\mathbf{c})=\wt(\mathbf{c}_{1})=i$ $(d \leqslant i \leqslant 7)$, Supp$(\mathbf{c})$=Supp$(\mathbf{c}_{1})$, but $\mathbf{c}$ is not a multiple of $\mathbf{c}_{1}$. Then we can find a nonzero element $\beta \in$ GF$(q)$ such that $\mathbf{c}_{2}=\mathbf{c}-\beta \mathbf{c}_{1} \in \mathcal{C}$, Supp$(\mathbf{c}_{2})\subseteq \Supp(\mathbf{c})$ and $\wt(\mathbf{c}_{2})\leqslant i-1$. This clearly derives a contradiction if $i=d$. So we let $i\geqslant d+1$ next.

If $i=6$ then we only need to let $m$ be even by Lemma \ref{l4.1}. Clearly $\wt(\mathbf{c}_{2})=5$ as the minimum distance $d=5$. By Lemma \ref{l4,3} (ii), Supp$(\mathbf{c}_{2})\in {\cal B}_{\sigma_{5,2},q+1}$. Since Supp$(\mathbf{c}_{2})\subseteq \Supp(\mathbf{c})$, one has  Supp$(\mathbf{c}) \in \mathcal{B}_{\sigma_{6,3},q+1}^{0}$.  But this contradicts $\mathcal{B}_{6}(\mathcal{C})=\mathcal{B}_{\sigma_{6,3},q+1} \setminus \mathcal{B}_{\sigma_{6,3},q+1}^{0}$.

Now we deal with the only remaining case of $i=7$. One has  $\wt(\mathbf{c}_{2})=5,6$  and if $\wt(\mathbf{c}_{2})=5$ then $m$ must be even  by Lemma \ref{l4.1}. If $\wt(\mathbf{c}_{2})=6$, then from Lemma \ref{l4,3} (i) and (ii), $\sigma_{6,3}$(Supp$(\mathbf{c}_{2}))=0$.  If $\wt(\mathbf{c}_{2})=5$ and $m$ is even, then  $\sigma_{5,2}$(Supp$(\mathbf{c}_{2}))=0$  by Lemma \ref{l4,3} (ii). Thus $\sigma_{6,3}(B)=0$ for any 6-subset $B$ with Supp$(\mathbf{c}_{2})\subseteq B\subseteq $ Supp$(\mathbf{c})$ by Lemma \ref{l3.3}. As a result, in either case, according to Lemma \ref{73u}, one has Supp$(\mathbf{c}) \in \mathcal{B}_{\sigma_{7,3},q+1}^{u}$, which is contrary to Lemma \ref{l4,3} (iii).  This completes the proof.\qed

We have the weight distribution formula for A$^{s}$MDS codes as follows.

\begin{lemma}{\rm\cite[Theorem 9]{1997Codes}}
Let $\mathcal{C}$ be an $[n,k,d]$ A$^{s}$MDS code over $\GF(q)$ with $s\geqslant 1$ and let the dual code $\mathcal{C}^{\bot}$ be an $[n,k,d^{\bot}]$ A$^{s^{\bot}}$MDS code.  Then the weight distribution $(A_{0}, A_{1}, \ldots, A_{n})$ of $\mathcal{C}$ satisfies
\begin{equation}\label{w1}
\begin{split}
A_{n-d^{\bot}+r}= \sum\limits^{n-d}_{j=d^{\bot}}\binom{j}{d^{\bot}-r}\bigg(\sum\limits^{j}_{i=d^{\bot}}(-1)^{i-d^{\bot}+r}\binom{j-d^{\bot}+r}{j-i}\bigg)A_{n-j}\\
+\binom{n}{d^{\bot}-r}\sum\limits^{r-1}_{i=0}(-1)^{i}\binom{n-d^{\bot}+r}{i}(q^{k-d^{\bot}+r-i}-1)
\end{split}
\end{equation}
for $r=1,2,\ldots, d^{\bot}.$ In particular, $A_{d},\ldots,A_{n-d^{\bot}}$ determine the weight distribution of $\mathcal{C}$ completely.
\end{lemma}

In particular, we have the following weight distribution formula for near MDS codes.

\begin{lemma}{\rm\cite[Theorem 4.1]{Stefan1995On}}
Let $\mathcal{C}$ be an $[n, k, n-k]$ near MDS code over $\GF(q)$. Then the weight distribution $(A_{0}, A_{1}, \ldots, A_{n})$ of $\mathcal{C}$ is given by
\begin{equation}\label{w2}
A_{n-k+s}=\binom{n}{k-s}\sum\limits_{j=0}^{s-1}(-1)^{j}\binom{n-k+s}{j}(q^{s-j}-1)+(-1)^{s}\binom{k}{s}A_{n-k}
\end{equation}
for $s\in \{1,2,\ldots,k\}.$
\end{lemma}

\begin{lemma}\label{t4.5}
For $q=2^m$ with odd $m \geqslant 5$, the incidence structure $$(\mathcal{P}(\mathcal{C}_{(q,q+1,4,1)}),\mathcal{B}_{7}(\mathcal{C}_{(q,q+1,4,1)}))$$
is isomorphic to the complementary design of $(U_{q+1},\mathcal{B}^{u}_{\sigma_{7,3},q+1})$ with block set $\mathcal{\overline{B}}^{u}_{\sigma_{7,3},q+1}$.
\end{lemma}

\proof From Lemma \ref{l4,3},  if we index the coordinates of the codewords in $\mathcal{C}:=\mathcal{C}_{(q,q+1,4,1)}$ with the elements in $U_{q+1}$, then we may let $\mathcal{B}_{6}(\mathcal{C})=\mathcal{B}_{\sigma_{6,3},q+1}$ and $\mathcal{B}_{7}(\mathcal{C})\subseteq \mathcal{\overline{B}}^{u}_{\sigma_{7,3},q+1}$. So we only need to prove that
$$|\mathcal{B}_{7}(\mathcal{C})|=|\mathcal{\overline{B}}^{u}_{\sigma_{7,3},q+1}|.$$
 Since $\mathcal{C}_{(q,q+1,4,1)}$ is a near MDS code with parameters $[q+1,q-5,6]$ if $m$ is odd by Lemma \ref{l4.1}, according to Eq. (\ref{w2}) one has
$$A_{7}=-(q-5)A_{6}+\binom{q+1}{7}(q-1).$$
From Lemmas \ref{4.4}  and  \ref{lemma 24},  $$A_{6}=|\mathcal{B}_{6}(\mathcal{C})|(q-1)=\frac{(q-8)(q-1)}{30}\binom{q+1}{4}.$$ Then
$$\binom{q+1}{7}-|\mathcal{B}_{7}(\mathcal{C})|=\binom{q+1}{7}-\frac{A_{7}}{q-1}=\frac{(q-5)(q-8)}{30}\binom{q+1}{4},$$
which is the same as $|\mathcal{B}^{u}_{\sigma_{7,3},q+1}|$ for odd $m$ from (\ref{e152}). This completes the proof.\qed

\begin{theorem}\label{t4.6}
For $q=2^m$ with odd $m \geqslant 5$, the codewords of weight $7$ in $\mathcal{C}_{(q,q+1,4,1)}$ support a $4$-$(q+1,7,\lambda)$ design where $$\lambda=\binom{q-3}{3}-\frac{7(q-5)(q-8)}{6}.$$
\end{theorem}

\proof The desired conclusion follows from Lemma \ref{t4.5} and Eq. (\ref{e3}). \qed

\begin{lemma}\label{t4.7}
For $q=2^m$ with even $m \geqslant 4$, the incidence structure $$(\mathcal{P}(\mathcal{C}_{(q,q+1,4,1)}),\mathcal{B}_{7}(\mathcal{C}_{(q,q+1,4,1)}))$$
is isomorphic to the complementary design of $(U_{q+1},\mathcal{B}^{u}_{\sigma_{7,3},q+1})$ with block set $\mathcal{\overline{B}}^{u}_{\sigma_{7,3},q+1}$.
\end{lemma}

\proof Similarly to the proof of Lemma \ref{t4.5}, we only need to prove that $|\mathcal{B}_{7}(\mathcal{C}_{(q,q+1,4,1)})|=| \mathcal{\overline{B}}^{u}_{\sigma_{7,3},q+1}|$ by Lemma \ref{l4,3}. Since $\mathcal{C}_{(q,q+1,4,1)}$ is an A$^{2}$MDS code with parameters $[q+1,q-5,5]$ if $m$ is even by Lemma \ref{l4.1}, then according to Eq. (\ref{w1}) one has
$$A_{7}=\binom{q+1}{7}(q-1)-(q-5)A_{6}-\frac{(q-4)(q-5)}{2}A_{5}.$$
From Lemmas \ref{4.4}, \ref{lemma23} and \ref{l10}, one has
$$\left\{
    \begin{array}{l}
A_{6}=|\mathcal{B}_{6}(\mathcal{C}_{(q,q+1,4,1)})|(q-1)=\frac{(q-4)(q-16)(q-1)}{120}\binom{q+1}{3},       \\
A_{5}=|\mathcal{B}_{5}(\mathcal{C}_{(q,q+1,4,1)})|(q-1)=\frac{(q-1)}{10}\binom{q+1}{3}.       \\
    \end{array}
  \right.
$$
Then $$\binom{q+1}{7}-|\mathcal{B}_{7}(\mathcal{C}_{(q,q+1,4,1)})|=\binom{q+1}{7}-\frac{A_{7}}{q-1}=  \frac{(q-4)(q-5)(q-10)}{120}\binom{q+1}{3},$$
which is the same as $|\mathcal{B}^{u}_{\sigma_{7,3},q+1}|$ for even $m$ from (\ref{e152}). This completes the proof. \qed

\begin{theorem}\label{t4.8}
For $q=2^m$ with even $m \geqslant 4$, the codewords of weight $7$ in $\mathcal{C}_{(q,q+1,4,1)}$ support a $3$-$(q+1,7,\lambda)$ design where $$\lambda=\binom{q-2}{4}-\frac{7(q-4)(q-5)(q-10)}{24}.$$
\end{theorem}

\proof The desired conclusion follows from Lemma \ref{t4.7} and Eq. (\ref{e3}). \qed

Denote by $\mathcal{C}$ the narrow-sense BCH code $\mathcal{C}_{(q,q+1,4,1)}$ over GF$(q)$. From what has been investigated  up to now, we have Tables 1 and 2 for $t$-designs supported by the codewords of a fixed weight in ${\cal C}$.

\begin{table}[h]
\centering
\begin{tabular}{c|c}
    \hline
   Block sets & \text{Designs} \\
   \hline
     $\mathcal{B}_{5}(\mathcal{C})\cong\mathcal{B}_{\sigma_{5,2},q+1}$  & $3$-$(q+1,5,1)$  \\

    $\mathcal{B}_{6}(\mathcal{C})\cong\mathcal{B}_{\sigma_{6,3},q+1} \setminus \mathcal{B}^{0}_{\sigma_{6,3},q+1}$  & $3$-$(q+1,6,\frac{(q-4)(q-16)}{6})$ \\
   $\mathcal{B}_{7}(\mathcal{C})\cong\binom{U_{q+1}}{7} \setminus \mathcal{B}^{u}_{\sigma_{7,3},q+1}$  & $3$-$(q+1,7,\binom{q-2}{4}-\frac{7(q-4)(q-5)(q-10)}{24})$ \\
    \hline
  \end{tabular}
  \caption{Known designs supported by $\mathcal{C}_{(q,q+1,4,1)}$ with $q=2^{2s}$}
  \end{table}

\begin{table}[h]
\centering
\begin{tabular}{c|c}
    \hline
   Block sets & \text{Designs} \\
   \hline
    $\mathcal{B}_{6}(\mathcal{C})\cong\mathcal{B}_{\sigma_{6,3},q+1}$  & $4$-$(q+1,6,\frac{q-8}{2})$ \\
   $\mathcal{B}_{7}(\mathcal{C})\cong\binom{U_{q+1}}{7} \setminus \mathcal{B}^{u}_{\sigma_{7,3},q+1}$  & $4$-$(q+1,7,\binom{q-3}{3}-\frac{7(q-5)(q-8)}{6})$\\
    \hline
  \end{tabular}
\caption{Known designs supported by $\mathcal{C}_{(q,q+1,4,1)}$ with $q=2^{2s+1}$}
\end{table}

According to Magma \cite{Cannon2005Handbook} experiments, we have the following two examples.

\begin{example}
Let $q=2^6$. Then $\mathcal{C}_{(64,65,4,1)}$ has parameters $[65,59,5]$ and its weight distribution equals

$1+275184z^5+66044160z^6+39476324160z^7+18256982332680z^8+7271676138046320z^9+2565751348965796992z^{10}+\ldots+A_{i}z^i+\ldots+A_{65}z^{65}.$

The codewords of weight $5$  support a $3$-$(65,5,1)$ design, and the codewords of weight $6$  support a $3$-$(65,6,480)$ design. Moreover, the codewords of weight $7$  support a $3$-$(65,7,502090)$ design.
\end{example}

\begin{example}
Let $q=2^5$. Then $\mathcal{C}_{(32,33,4,1)}$ has parameters $[33,27,6]$ and its weight distribution equals

$1+1014816z^6+105033456z^7+11116421316z^8+948713422800z^9+70662246969600z^{10}+\ldots+A_{i}z^i+\ldots+A_{33}z^{33}.$

The codewords of weight $6$ support a $4$-$(33,6,12)$ design, and the codewords of weight $7$  support a $4$-$(33,7,2898)$ design.
\end{example}

\begin{problem}
Do the codewords of a given weight $k>7$ in $\mathcal{C}_{(q,q+1,4,1)}$ support a $t$-$(q+1,k,\lambda)$ design$?$ If yes, determine the parameters $t$ and $\lambda$.
\end{problem}

\section{Trace codes supporting $4$-designs}
This section is divided into three subsections. In subsection 5.1 we  define  trace codes and use $\Tr_{q^2/q}(\mathcal{C}_{\{1,2,3\}})$ to represent the dual code $\mathcal{C}^\bot_{(q,q+1,4,1)}$. We show that the trace code $\Tr_{q^2/q}(\mathcal{C}_{\{1,2,3\}})$  supports the supplementary design of $(U_{q+1},\mathcal{B}_{\sigma_{5,3},q+1}^{b})$ with parameters $4$-$(q+1,5,5)$ where $q=2^{2s+1}$. In subsection 5.2 we prove that the set of supports of a fixed weight in $\Tr_{q^2/q}(\mathcal{C}_{\{1,2,3\}})$ is invariant under the action of a $3$-transitive group, which is isomorphic to $\PGL(2,q)$. In the last subsection we employ the known information on the permutation character of $\PGL(2,q)$ to prove that the $4$-$(q+1,5,5)$ designs produced from $\Tr_{q^2/q}(\mathcal{C}_{\{1,2,3\}})$ are isomorphic to the designs with the same parameters constructed by Alltop \cite{1969An} in 1969.

\subsection{Tr$_{q^2/q}(\mathcal{C}_{\{1,2,3\}})$ and $4$-$(q+1,5,5)$ design}
Let $\mathcal{C}$ be a code of length $n$ over GF($q^m$). Then we call $\mathcal{C}\cap \text{GF}(q)^{n}$  the {\em subfield subcode} over $\GF(q)$ and usually denote it by $\mathcal{C}|_{\text{GF}(q)}$. The {\em trace code} of $\mathcal{C}$ is  defined by
$$ \Tr_{q^m/q}(\mathcal{C})=\big\{\big(\Tr_{q^m/q}(c_{0}),\Tr_{q^m/q}(c_{1})\ldots, \Tr_{q^m/q}(c_{n-1})\big):(c_{0},c_{1},\ldots,c_{n-1})\in \mathcal{C}\big\},
$$
where $\Tr_{q^m/q}$ represents the trace function from GF$(q^m)$ to GF$(q)$.  Delsarte \cite{Delsarte1975On} stated that
$$
(\Tr_{q^m/q}(\mathcal{C}))^{\bot}=\mathcal{C}^{\bot}|_{\text{GF}(q)}.
$$

Ding et al. \rm{\cite{2020The}} gave a cyclic code over GF$(q^2)$ of length $q+1$ by defining
$$\mathcal{C}_{\{3,5\}}=\big\{(a_{3}u^{3}+a_{q-2}u^{q-2}+a_{5}u^{5}+a_{q-4}u^{q-4})_{u\in U_{q+1}}: a_{3},a_{5},a_{q-2},a_{q-4}\in \text{GF}(q^2)\big\}.$$
Thus the trace code of $\mathcal{C}_{\{3,5\}}$ is given by
$$\Tr_{q^{2}/q}(\mathcal{C}_{\{3,5\}})=\{(\Tr_{q^{2}/q}(au^3+bu^{5}))_{u\in U_{q+1}}: a,b\in \text{GF}(q^{2})\}.$$
Similarly, we define another cyclic code and  its trace code by
$$\mathcal{C}_{\{1,2,3\}}=\left \{ \begin{array}{cl}
 (a_{1}u+a_{q}u^{q}+a_{2}u^{2}+a_{q-1}u^{q-1}+a_{3}u^{3}+a_{q-2}u^{q-2})_{u\in U_{q+1}}:
 \\  \qquad\qquad \qquad \qquad \qquad \qquad a_{1},a_{2},a_{3},a_{q},a_{q-1},a_{q-2}\in \GF(q^2)\end{array}
                  \right\},$$
\begin{equation}\label{e16}
\Tr_{q^{2}/q}(\mathcal{C}_{\{1,2,3\}})=\{(\Tr_{q^{2}/q}(au+bu^{2}+cu^{3}))_{u\in U_{q+1}}: a,b,c\in \text{GF}(q^{2})\}.
\end{equation}
It is immediate that if $(c_{u})_{u\in U_{q+1}}\in \mathcal{C}_{\{1,2,3\}}$, then $(c^{q}_{u})_{u\in  U_{q+1}}\in \mathcal{C}_{\{1,2,3\}}$. From \cite[Lemma 7]{2010Galois} we derive that
\begin{equation}\label{e15}
\Tr_{q^2/q}(\mathcal{C}_{\{1,2,3\}})=\mathcal{C}_{\{1,2,3\}}|_{\GF(q)}
\end{equation}
and
$$\Tr_{q^2/q}(\mathcal{C}_{\{1,2,3\}}^{\bot})=\mathcal{C}_{\{1,2,3\}}^{\bot}|_{\GF(q)}.$$

 Recall the proof of \rm{\cite[Theorem 34]{Tang2020An}}, where the trace expression of $\mathcal{C}_{(q,q+1,4,1)}^{\bot}$ was given. Next combining  (\ref{e16}) and (\ref{e15}) yields that we may identify the codes $\mathcal{C}_{(q,q+1,4,1)}^{\bot}$,  $\Tr_{q^{2}/q}(\mathcal{C}_{\{1,2,3\}})$ and $\mathcal{C}_{\{1,2,3\}}|_{\GF(q)}$; and they have parameters $[q+1,6,q-5]$ by Lemma \ref{l4.1}.

\begin{lemma}\label{l5.3}
Let $f(u)=\Tr_{q^{2}/q}(au+bu^{2}+cu^{3})$ where $(a,b,c)\in \GF(q^{2})^{3}\backslash \{\mathbf{0}\}$. Define  $\zero(f)$\\ $=\{u\in U_{q+1}: f(u)=0\}.$ Then $|\zero(f)|\leqslant 6$ and we have the following.

$(1)$ $|\zero(f)|=6$ if and only if $a=\frac{\tau\sigma_{6,2}(B)}{\sqrt{\sigma_{6,6}(B)}}, b=\frac{\tau\sigma_{6,1}(B)}{\sqrt{\sigma_{6,6}(B)}}$ and $c=\frac{\tau}{\sqrt{\sigma_{6,6}(B)}}$, where $B\in \mathcal{B}_{\sigma_{6,3},q+1}$ and $\tau \in$ $\GF(q)^{*}$.

$(2)$ $|\zero(f)|=5$ if and only if there is $B\in \mathcal{B}^{b}_{\sigma_{5,3},q+1}$ and $u_{i}\in B$ such that $a=\frac{\tau(\sigma_{5,2}(B)+u_{i}\sigma_{5,1}(B))}{\sqrt{u_{i}\sigma_{5,5}(B)}}$,
$b=\frac{\tau(\sigma_{5,1}(B)+u_{i})}{\sqrt{u_{i}\sigma_{5,5}(B)}}$ and $c=\frac{\tau}{\sqrt{u_{i}\sigma_{5,5}(B)}}$, where $\tau \in$ $\GF(q)^{*}$.
\end{lemma}

\proof When $u\in U_{q+1}$ one has
$$ f(u) =au+bu^2+cu^3+a^qu^{-1}+b^qu^{-2}+c^qu^{-3}. $$
If $c=0$, then $f(u)=au+bu^2+a^qu^{-1}+b^qu^{-2}=\frac{1}{u^2}(bu^4+au^3+a^qu+b^q)$. This shows that $\zero(f)\leqslant 4$. Next let $c\ne 0$. Then,
\begin{equation}\label{e18}
f(u)=\frac{1}{u^{3}}(cu^{6}+bu^{5}+au^{4}+a^{q}u^{2}+b^{q}u+c^{q}).
\end{equation}
Hence, $|\text{zero}(f)|\leqslant 6$.

(1) The proof can be found in \rm{\cite[Lemma 33]{Tang2020An}}.

(2) Assume that $|\text{zero}(f)|=5$ and let $B=\{u_{1},u_{2},\dots,u_{5}\}\in \binom{U_{q+1}}{5}$ be the set of the five roots of $f$ in $U_{q+1}$. Then we have another root $\alpha$ in the splitting field of $f$. So $f(u)=\frac{c(u+\alpha)\prod^{5}_{j=1}(u+u_{j})}{u^{3}}$. By Vieta's formula, $c\alpha\sigma_{5,5}(B)=c^q$, $c(\sigma_{5,5}(B)+\alpha\sigma_{5,4}(B))=b^q$, $c(\sigma_{5,4}(B)+\alpha\sigma_{5,3}(B))=a^q$, $c(\sigma_{5,3}(B)+\alpha\sigma_{5,2}(B))=0$, $c(\sigma_{5,2}(B)+\alpha\sigma_{5,1}(B))=a$, $c(\sigma_{5,1}(B)+\alpha)=b$. One obtains $\alpha=\frac{c^{q-1}}{\sigma_{5,5}(B)}$ from $\alpha\sigma_{5,5}(B)=c^{q-1}$. Note that $\alpha^{q+1}=\frac{c^{(q-1)(q+1)}}{\sigma_{5,5}(B)^{q+1}}=1$ as $\sigma_{5,5}(B)\in U_{q+1}$. Then $\alpha\in U_{q+1}$. This shows that $\alpha=u_{i}$ is a double root of $f$ in $U_{q+1}$ for some $u_{i}\in B$. One also obtains $B\in \mathcal{B}^{b}_{\sigma_{5,3},q+1}$ from $\sigma_{5,3}(B)+u_{i}\sigma_{5,2}(B)=0$ and $c=\frac{\tau}{\sqrt{u_{i}\sigma_{5,5}(B)}}$ from $u_{i}\sigma_{5,5}(B)=c^{q-1}$, where $\tau \in$ GF$(q)^{*}$. Then $a=\frac{\tau(\sigma_{5,2}(B)+u_{i}\sigma_{5,1}(B))}
{\sqrt{u_{i}\sigma_{5,5}(B)}}$ and $b=\frac{\tau(\sigma_{5,1}(B)+u_{i})}{\sqrt{u_{i}\sigma_{5,5}(B)}}$.

Conversely, assume that $B=\{u_{1},u_{2},\ldots,u_{5}\}\in \mathcal{B}^{b}_{\sigma_{5,3},q+1}$ and there is $u_{i}\in B$ such that $a=\frac{\tau(\sigma_{5,2}(B)+u_{i}\sigma_{5,1}(B))}{\sqrt{u_{i}\sigma_{5,5}(B)}}$,
$b=\frac{\tau(\sigma_{5,1}(B)+u_{i})}{\sqrt{u_{i}\sigma_{5,5}(B)}}$ and $c=\frac{\tau}{\sqrt{u_{i}\sigma_{5,5}(B)}}$, where $\tau \in$ GF$(q)^{*}$. Then $f(u)=\frac{c(u+u_{i})\prod^{5}_{i=1}(u+u_{j})}{u^{3}}$. Thus, zero$(f)=B$ and $|\text{zero}(f)|=5$. \qed

\begin{theorem}\label{t5.4}
Let $q=2^m$ with odd $m \geqslant 5$. Then the incidence structure $$\mathbb{D}=(U_{q+1},\mathcal{B}_{q-4}(\Tr_{q^{2}/q}(\mathcal{C}_{\{1,2,3\}})))$$
forms a $4$-$(q+1,q-4,\binom{q-4}{4})$ design and its supplementary design is isomorphic to $(U_{q+1},\mathcal{B}^{b}_{\sigma_{5,3},q+1})$, which forms a $4$-$(q+1,5,5)$ design.
\end{theorem}

\proof Lemma \ref{l5.3} (2) shows that the incidence structure $(U_{q+1},\mathcal{B}_{q-4}(\Tr_{q^{2}/q}(\mathcal{C}_{\{1,2,3\}})))$
is isomorphic to the supplementary design of $(U_{q+1},\mathcal{B}^{b}_{\sigma_{5,3},q+1})$, which is a $4$-$(q+1,5,5)$ design from Theorem \ref{t3.4} (1). According to Eq. (\ref{e2}), $(U_{q+1},\mathcal{B}_{q-4}(\Tr_{q^{2}/q}(\mathcal{C}_{\{1,2,3\}})))$ forms a $4$-$(q+1,q-4,\lambda)$ design, where
$$\lambda=5\frac{\binom{q+1-4}{5}}{\binom{q+1-4}{5-4}}=\binom{q-4}{4}.
$$
This completes the proof. \qed

According to Magma \cite{Cannon2005Handbook} experiments, we have the following example.
\begin{example}
Let $q=2^5$. Then the trace code $\Tr_{q^{2}/q}(\mathcal{C}_{\{1,2,3\}})$ has parameters $[33,6,27]$ and its weight distribution equals

$1+1014816z^{27}+1268520z^{28}+20296320z^{29}+64609952z^{30}+210132384z^{31}+399584823z^{32}+376835008z^{33}.$

The codewords of weight $27$  support a $4$-$(33,27,14040)$ design, and the codewords of weight $28$ support a $4$-$(33,28,20475)$ design.
\end{example}

\begin{lemma}\label{l5.1}
For $q=2^m$ with even $m \geqslant 4$, one has $\mathcal{B}_{\sigma_{4,2},q+1}^{u}=\mathcal{B}_{\sigma_{4,2}^{2}+\sigma_{4,1}\sigma_{4,3},q+1}$ $($see $(\ref{Bf})$ for definition of $\mathcal{B}_{f,q+1})$.
\end{lemma}

\proof From the proof of Theorem \ref{t3.2}, $\mathcal{B}_{\sigma_{4,2},q+1}^{u}=\Big\{B\in \binom{U_{q+1}}{4}: \sigma_{5,2}(B\cup \{a\})=0\ \text{for some}\ a\in U_{q+1}\setminus B\Big\}$. Combine \cite[Lemmas 12, 18]{Tang2020An} and \cite[Lemma 18]{2020The} to yield that $\mathcal{B}_{\sigma_{4,2},q+1}^{u}=\mathcal{B}_{\sigma_{4,2}^{2}+\sigma_{4,1}\sigma_{4,3},q+1}$. \qed

\begin{corollary}
Let $q=2^m$ and $m \geqslant 4$ be even. Then  $$(U_{q+1},\mathcal{B}_{4}((\Tr_{q^m/q}(\mathcal{C}_{\{3,5\}}))^{\bot}))$$
from the minimum weight codewords in $(\Tr_{q^m/q}(\mathcal{C}_{\{3,5\}}))^{\bot}$ is isomorphic to $(U_{q+1},\mathcal{B}^{u}_{\sigma_{4,2},q+1})$, which is a $3$-$(q+1,4,2)$ design.
\end{corollary}

\proof  From  \cite[Theorem 27]{2020The}, one has that the supports of all codewords of the minimum weight in $\mathcal{C}_{\{3,5\}}^{\bot}|_{\GF(q)}$ support a design
 isomorphic to $(U_{q+1},\mathcal{B}_{\sigma_{4,2}^{2}+\sigma_{4,1}\sigma_{4,3},q+1})$. Then the conclusion can be obtained directly from Lemma \ref{l5.1} and Theorem \ref{t3.2}, noticing $(\Tr_{q^2/q}(\mathcal{C}_{\{3,5\}}))^{\bot}=\mathcal{C}_{\{3,5\}}^{\bot}|_{\GF(q)}$ from the proof of \cite[Theorem 21]{2020The}. \qed

\subsection{Automorphism group of the set of supports in $\Tr_{q^{2}/q}(\mathcal{C}_{\{1,2,3\}})$}
Let $\PGL(2,q)$ denote the projective general linear group acting on the points of the projective line PG$(1,q)$. We may write  any vector in the $(q+1)$-dimensional vector space $\GF(q)^{q+1}$  as $(c_{x})_{x\in \PG(1,q)}$, where $c_{x}\in \GF(q)$. There is an induced action of $\PGL(2,q)$ on $\GF(q)^{q+1}$ defined by the translation
$$\pi:(c_{x})_{x\in \PG(1,q)} \mapsto (c_{\pi(x)})_{x\in \PG(1,q)},$$
where $(c_{x})_{x\in \PG(1,q)}\in \GF(q)^{q+1}$ and $\pi\in \PGL(2,q)$. The set $\mathcal{B}_{k}(\mathcal{C})$ of supports of all codewords of weight $k$ is said to be {\em invariant} under $\PGL(2,q)$ if Supp$((c_{\pi(x)})_{x\in \PG(1,q)})\in\mathcal{B}_{k}(\mathcal{C})$ for every permutation $\pi\in\PGL(2,q)$ and any codeword $(c_{x})_{x\in\PG(1,q)}$ of weight $k$ in $\mathcal{C}$.

Let Stab$_{U_{q+1}}=\{g\in\PGL(2,q^2):g(U_{q+1})=U_{q+1}\}$ be the {\em setwise stabilizer} of $U_{q+1}$ under the action of $\PGL(2,q^2)$ on  PG$(1,q^2)$. Then the following lemma presents the specific structure of Stab$_{U_{q+1}}$.

\begin{lemma}{\rm\cite[Corollary 6]{2020The}}\label{l5.62}
Let $q=2^m$. Then $\stab_{U_{q+1}}$ is generated by  three types of linear fractional transformations as follows.
\begin{enumerate}
  \item [(1)] $u\mapsto u_{0}u$, where $u_{0}\in U_{q+1}$;
  \item [(2)] $u \mapsto u^{-1}$;
  \item [(3)] $u\mapsto \frac{u+c^q}{cu+1}$, where $c\in $GF$(q^2)^{*} \setminus U_{q+1}$.
\end{enumerate}
\end{lemma}

\begin{proposition}{\rm\cite[Proposition 7]{2020The}}\label{p5.7}
Let $q=2^m$ and $\stab_{U_{q+1}}$ the setwise stabilizer of $U_{q+1}$. Then $\stab_{U_{q+1}}$ is conjugate in $\PGL(2,q^2)$ to the group $\PGL(2,q)$ and its action on $U_{q+1}$ is equivalent to the action of $\PGL(2,q)$ on $\PG(1,q)$.
\end{proposition}

We will establish the invariance of the  support set of  any fixed weight in $\Tr_{q^{2}/q}(\mathcal{C}_{\{1,2,3\}})$ under the action of Stab$_{U_{q+1}}$ in the following theorem.

\begin{theorem}\label{t5.7}
Let $q=2^{m}$ with $m\geqslant 4$. Let $k$ be a positive integer with $k \leqslant q+1$ and $A_{k}(\Tr_{q^{2}/q}(\mathcal{C}_{\{1,2,3\}}))>0$. Then $\mathcal{B}_{k}(\Tr_{q^{2}/q}(\mathcal{C}_{\{1,2,3\}}))$ is invariant under the action of $\stab_{U_{q+1}}$ and hence it forms the block set of a $3$-$(q+1,k,\lambda)$ design for some positive integer $\lambda$ when $3<k<q+1$.
\end{theorem}

\proof The proof is similar to that of \cite[Theorem 23]{2020The}. In view of Proposition \ref{p5.7}, we only need to prove the former part of the conclusion because it is well-known that $\PGL(2,q)$ acts on $\PG(1,q)$ 3-transitively.  As a result, it suffices to show that if $\mathbf{c}\in \Tr_{q^{2}/q}(\mathcal{C}_{\{1,2,3\}})$ and $\pi$ is a  transformation in Lemma \ref{l5.62}, then there exists a  $\mathbf{c}_{1}\in \Tr_{q^{2}/q}(\mathcal{C}_{\{1,2,3\}})$ such that Supp$(\pi(\mathbf{c}))$= Supp$(\mathbf{c}_{1})$. Let $\mathbf{c}(a,b,d)$ denote the codeword $(\Tr_{q^{2}/q}(au+bu^{2}+du^{3}))_{u\in U_{q+1}}$ of $\Tr_{q^{2}/q}(\mathcal{C}_{\{1,2,3\}})$, where $a,b,d\in \text{GF}(q^{2})$.  The following three cases for $\pi$ will be handled.

If $\pi:u\mapsto u_{0}u$ for some $u_{0}\in U_{q+1}$, then it is obvious that $\pi(\mathbf{c}(a,b,d))=\mathbf{c}(au_{0},bu_{0}^{2},du_{0}^{3})$. Hence Supp$(\pi(\mathbf{c}(a,b,d)))$ = Supp$(\mathbf{c}(au_{0},bu_{0}^{2},du_{0}^{3}))$.

If $\pi:u \mapsto u^{-1}$, then obviously $\pi(\mathbf{c}(a,b,d))=\mathbf{c}(a^q,b^q,d^q)$. Thus Supp$(\pi(\mathbf{c}(a,b,d)))$ =Supp$(\mathbf{c}(a^q,b^q,d^q))$.

Finally let $\pi:u \mapsto \frac{u+c^{q}}{cu+1}$ where $c \in \GF(q^{2})^{*}\setminus U_{q+1}$. Let $f(u) = \Tr_{q^{2}/q}(au+bu^{2}+du^{3})$ and $A = cu +1$. Then $u +c^{q} = uA^{q}$. Simple computation gives
\begin{equation}\label{e30}
\left.
    \begin{array}{l}
      f(\frac{u+c^{q}}{cu+1})\\
      = \Tr_{q^{2}/q}(a(\frac{u+c^{q}}{cu+1})+b(\frac{u+c^{q}}{cu+1})^{2}+d(\frac{u+c^{q}}{cu+1})^{3})\\
       =\Tr_{q^{2}/q}(\frac{a(u+c^{q})(cu+1)^{2}+b(u+c^{q})^{2}(cu+1)+d(u+c^{q})^{3}}{(cu+1)^{3}})\\
     = \Tr_{q^{2}/q}(\frac{auA^{q+2}+bu^{2}A^{2q+1}+du^{3}A^{3q}}{A^{3}}) \\
     =\frac{auA^{q+2}+bu^{2}A^{2q+1}+du^{3}A^{3q}}{A^{3}}+ \frac{a^{q}u^{q}A^{2q+1}+b^{q}u^{2q}A^{q+2}+d^{q}u^{3q}A^{3}}{A^{3q}} \\
     = \frac{auA^{4q+2}+bu^{2}A^{5q+1}+du^{3}A^{6q}+a^{q}u^{q}A^{2q+4}+b^{q}u^{2q}A^{q+5}+d^{q}u^{3q}A^{6}}{A^{3}A^{3q}} \\
      = \frac{(auA^{4q+2}+bu^{2}A^{5q+1}+du^{3}A^{6q})+(auA^{4q+2}+bu^{2}A^{5q+1}+du^{3}A^{6q})^{q}}{A^{3}A^{3q}}\\
     = \frac{1}{A^{3}A^{3q}}\Tr_{q^2/q}(A^{4q+2}au+A^{5q+1}bu^{2}+A^{6q}du^{3}),\\
    \end{array}
  \right.
\end{equation}
and
\begin{equation}\label{e20}
\left.
  \begin{array}{l}
   auA^{4q+2} \\=auA^{4q}A^{2}  \\
   =au(cu+1)^{4q}(cu+1)^{2}\\
     =au(c^{4q}u^{4q}+1)(c^{2}u^{2}+1) \\
   =a(u+c^{2}u^{3}+c^{4q+2}u^{-1}+c^{4q}u^{-3}).  \\
  \end{array}
\right.
\end{equation}
Similarly, we have
\begin{equation}\label{e21}
\left.
  \begin{array}{l}
   bu^{2}A^{5q+1}  \\=bu^{2}A^{4q}A^{q}A  \\
     =bu^{2}(cu+1)^{4q}(cu+1)^{q}(cu+1)  \\
     = bu^{2}(c^{4q}u^{4q}+1)(c^{q}u^{q}+1)(cu+1) \\
     = b(cu^{3}+(c^{q+1}+1)u^{2}+c^{q}u+c^{4q+1}u^{-1}+(c^{4q}+c^{5q+1})u^{-2}+c^{5q}u^{-3}),\\
  \end{array}
\right.
\end{equation}
and
\begin{equation}\label{e22}
\left.
  \begin{array}{l}
    du^{3}A^{6q} \\= du^{3}A^{4q}A^{2q} \\
     =du^{3}(cu+1)^{4q}(cu+1)^{2q}\\
     =du^{3}(c^{4q}u^{4q}+1)(c^{2q}u^{2q}+1)  \\
     =d(u^{3}+c^{2q}u+c^{4q}u^{-1}+c^{6q}u^{-3}).\\
  \end{array}
\right.
\end{equation}
Combining Eq. (\ref{e20})-(\ref{e22}) gives
\begin{equation}\label{e23}
\Tr_{q^2/q}(auA^{4q+2}+bu^{2}A^{5q+1}+du^{3}A^{6q})=\Tr_{q^2/q}(a'u+b'u^{2}+c'u^{3}),
\end{equation}
where
$$\left\{
    \begin{array}{l}
     a'=a+bc^{q}+dc^{2q}+a^{q}c^{2q+4}+b^{q}c^{q+4}+d^{q}c^{4} , \\
     b'=b(c^{q+1}+1)+b^{q}(c^{4}+c^{q+5}),\\
     d'=ac^{2}+bc+d+a^{q}c^{4}+b^{q}c^{5}+a^{q}c^{6}.\\
    \end{array}
  \right.
$$
Plugging (\ref{e23}) into (\ref{e30}) yields
$$f(\frac{u+c^{q}}{cu+1})=\frac{1}{A^{3}A^{3q}}\Tr_{q^2/q}(a'u+b'u^{2}+d'u^{3}).
$$

So we have Supp$(\pi(\mathbf{c}(a,b,d)))$=Supp$(\mathbf{c}(a',b',d'))$. The  conclusion then follows. \qed

\begin{corollary}\label{c5.92}
Let $q=2^{m}$ and $m\geqslant 4$ be odd. Then $\mathcal{B}^{b}_{\sigma_{5,3},q+1}$ is invariant under the action of  $\stab_{U_{q+1}}$.
\end{corollary}

\proof Theorem \ref{t5.7} implies that $\mathcal{B}_{q-4}(\Tr_{q^{2}/q}(\mathcal{C}_{\{1,2,3\}}))$ is invariant under the action of $\stab_{U_{q+1}}$ by noting $A_{q-4}(\mathcal{C}_{\{1,2,3\}}))>0$ from Eq. (\ref{w1}). According to Theorem \ref{t5.4}, we have
$$\mathcal{B}^{b}_{\sigma_{5,3},q+1}=\Bigg\{B\in \binom{U_{q+1}}{5}: (U_{q+1}\setminus B)\in \mathcal{B}_{q-4}(\Tr_{q^{2}/q}(\mathcal{C}_{\{1,2,3\}})) \Bigg\}.$$
Then the conclusion follows. \qed

\subsection{Relationship with Alltop's family}
The subgroup structure of $\PGL(2,q)$ is known \cite{1902Linear} and the permutation character $\chi$ for the action of $\PGL(2,q)$ on $\PG(1,q)$ is given in Table 3, where $\varphi(x)$ denotes Euler's totient.
\begin{table}[htpb]
\centering
\begin{tabular}{ccccc}
     \hline
Order of $g$ &   1  & 2 & $d|(q-1),d\neq 1$ &  $d|(q+1),d\neq 1$ \\

 Order of the centralizer of $g$ &  $q^3-q$ & $q$ & $q-1$ & $q+1$ \\
Number of conjugacy classes & 1 &  1 & $\varphi(d)/2$ & $\varphi(d)/2$\\
Number of fixed points   & $q+1$  & 1 & 2 & 0\\
   \hline
  \end{tabular}
\caption{Permutation character of PGL$(2,q)$, $q=2^m$}
\end{table}

Keranen and Kreher \cite{20043} gave complete information on the action of $\PGL(2,q)$ on all quadruples and quintuples of the projective line $\PG(1,q)$. In particular, whenever $q=2^m$ with $m$ being odd, under the action of $\PGL(2,q)$, all quintuples form exactly $\frac{q-2}{6}$ short orbits with stabilizer size $4$ and exactly $\frac{(q-2)(q-8)}{120}$ orbits with trivial stabilizer, see \cite[Theorem 3.3]{20043}. Alltop \cite{1969An} proved that the union of all those short orbits forms a $4$-$(q+1,5,5)$ design. In subsection 5.2 we showed that the
support set of all codewords with a given weight  in $\Tr_{q^{2}/q}(\mathcal{C}_{\{1,2,3\}})$ is invariant under the action of Stab$_{U_{q+1}}$ on $U_{q+1}$, which is equivalent to the action of $\PGL(2,q)$ on PG$(1,q)$.

Let supp$(\mathbb{D})$ be the supplementary design of the $4$-design $\mathbb{D}$ from Theorem \ref{t5.4}, which is isomorphic to the $4$-$(q+1,5,5)$ design corresponding to $(U_{q+1},\mathcal{B}_{\sigma_{5,3},q+1}^{b})$.  According to Magma \cite{Cannon2005Handbook} experiments, supp$({\mathbb{D}})$ is isomorphic to the Alltop's design with the same parameters when $q\in \{2^5,2^7\}$. In order to prove that supp$({\mathbb{D}})$ is isomorphic to the Alltop's design in general, we need the following lemmas.

\begin{lemma}\label{l5.11.2}
Let $q=2^m$ and $m$ be odd. Suppose that $B$ is a $5$-subset of $\PG(1,q)$ which is fixed by a non-identity element $f\in \PGL(2,q)$. Then we have the following.
\begin{enumerate}
  \item[$(1)$] f has order $2$; f has exactly one fixed point $\gamma$ and $\frac{q}{2}$ $2$-cycles $(\alpha_{i},f(\alpha_{i}))$, $1\leqslant i \leqslant \frac{q}{2}$.
  \item [$(2)$]$B=\{\alpha_{i},f(\alpha_{i}),\alpha_{j},f(\alpha_{j}),\gamma\}$ for some $1\leqslant i< j \leqslant \frac{q}{2}$.
\end{enumerate}
\end{lemma}

\proof Let ord$(f)=a$ and $B\in \binom{\PG(1,q)}{5}$. If $f(B)=B$, then it is not difficult to show that $B$ consists of $b\leqslant 2$ fixed points of $f$ and $c=\frac{5-b}{a}$ $a$-cycles of $f$. By Table 3, $b\leqslant 2$. Because $3\nmid q-1$, $5\nmid q+1$ when $m$ is odd, we must have $b=1$, $a=2$ and $c=2$. Then the conclusion follows.\qed

\begin{lemma}\label{l5.112} Let $q=2^m$ and $m$ be odd. 
Let $B\in \binom{U_{q+1}}{5}$ be a $5$-subset which is fixed by a non-identity element $f\in \stab_{U_{q+1}}$. Then there exists $h\in \stab_{U_{q+1}}$ and $\alpha\ne \beta\in U_{q+1}$ such that $B=h(\{\alpha,\frac{1}{\alpha},\beta,\frac{1}{\beta},1\})$.
\end{lemma}

\proof We apply Proposition \ref{p5.7}, Table 3 and Lemma \ref{l5.11.2}. Let $g: u\mapsto\frac{1}{u}$, $u\in U_{q+1}$. Then $g$ is an element of $\stab_{U_{q+1}}$ of order $2$. So $g$ has one fixed point $1$ and $\frac{q}{2}$ $2$-cycles $(\alpha_{i},g(\alpha_{i}))$, $1\leqslant i\leqslant \frac{q}{2}$ by Lemma \ref{l5.11.2} (1). Assume $B\in \binom{U_{q+1}}{5}$ is fixed by a non-identity element $f\in \stab_{U_{q+1}}$. According to Lemma \ref{l5.11.2} (1), ord$(f)=2$. By Table 3, there is only one conjugacy of order $2$ in $\stab_{U_{q+1}}$. Then there exists $h\in \stab_{U_{q+1}}$ such that $f=hgh^{-1}$. Since $B=f(B)$, we have
$B=hgh^{-1}(B).$
This shows that $h^{-1}(B)=g(h^{-1}(B))$, i.e., $h^{-1}(B)$ is fixed by $g$. From Lemma \ref{l5.11.2} (2), $h^{-1}(B)=\{\alpha,g(\alpha),\beta,g(\beta),1\}$ where  $(\alpha,g(\alpha))$ and $(\beta,g(\beta))$ are two  2-cycles in $g$.  As a result, $B=h(\{\alpha,\frac{1}{\alpha},\beta,\frac{1}{\beta},1\})$. \qed


\begin{lemma}\label{b455} For $q=2^m$ with odd $m \geqslant 5$,  $(U_{q+1},\mathcal{B}^{b}_{\sigma_{5,3},q+1})$ is isomorphic to the $4$-$(q+1,5,5)$ design constructed by Alltop \rm{\cite{1969An}}.
\end{lemma}

\proof Consider the action of Stab$_{U_{q+1}}$ on all $5$-subsets of $U_{q+1}$, which is  equivalent to the action of $\PGL(2,q)$ on  $5$-subsets of $\PG(1,q)$ by Proposition \ref{p5.7}. From Lemma \ref{l5.112}, any short orbit (under  the action of Stab$_{U_{q+1}}$) must have a quintuple representative $A=\{\alpha,\frac{1}{\alpha},\beta,\frac{1}{\beta},1\}$ for some $\alpha\ne \beta \in U_{q+1}\setminus \{1\}$. Clearly $\sigma_{5,3}(A)+\sigma_{5,2}(A)=0$, yielding $A\in \mathcal{B}^{b}_{\sigma_{5,3},q+1}$. By Corollary \ref{c5.92}, $\mathcal{B}_{\sigma_{5,3},q+1}^{b}$ is invariant under the action of $\stab_{U_{q+1}}$. As a consequence, we have $\mathcal{A}\subseteq \mathcal{B}_{\sigma_{5,3},q+1}^{b}$, where $\mathcal{A}$ is the set of all $5$-subsets in all short orbits. From \cite[Theorem 3.3]{20043}, we have $\frac{q-2}{6}$ short orbits whose stabilizers are all of size $4$. Hence $|\mathcal{A}|=\frac{q-2}{6}\cdot \frac{(q+1)q(q-1)}{4}=\binom{q+1}{4}$, which is the same as $|\mathcal{B}_{\sigma_{5,3},q+1}^{b}|$. So $\mathcal{B}^{b}_{\sigma_{5,3},q+1}$ equals the union of all short orbits and thus $(U_{q+1},\mathcal{B}^{b}_{\sigma_{5,3},q+1})$ is isomorphic to the $4$-$(q+1,5,5)$ design constructed by Alltop \cite{1969An}. \qed

\begin{theorem}
Let $q=2^m$ with odd $m \geqslant 5$. Then the incidence structure $$(U_{q+1},\mathcal{B}_{q-4}(\Tr_{q^{2}/q}(\mathcal{C}_{\{1,2,3\}})))$$
forms a $4$-$(q+1,q-4,\binom{q-4}{4})$ design and its supplementary design is isomorphic to the  $4$-$(q+1,5,5)$ design constructed by  Alltop \rm{\cite{1969An}}.
\end{theorem}

\proof The conclusion follows by combining Theorem \ref{t5.4} with Lemma \ref{b455}. \qed

Baartmans et al. \cite{Preparata} constructed  a class of 4-$(2^{2s}+ 1,5,2)$
designs by extending  the 3-designs formed by the minimum
weight codewords in the Preparata code of length $n = 2^{2s}$. Unfortunately, this class of 4-designs is not simple. In this paper we provide an infinite family of linear codes giving rise to an infinite family of simple
4-$(2^{2s+1}+ 1,5,5)$ designs; so far this class of designs has the smallest index among all known simple 4-$(q+1,5,\lambda)$  designs  derived from
 codes for prime powers $q$.

\section{Summary and concluding remarks}
Coding theory and design theory interact with each other intimately, with results and methods from one area being applied to the other. 
In this paper, we mainly investigated the topic of linear codes supporting $t=3,4$ designs by handling the incidence structures produced  from some variants of ESPs. Specifically, the main contributions of this paper are the following:
\begin{itemize}
  \item The codewords of weight $7$ in the BCH code $\mathcal{C}_{(q,q+1,4,1)}$ for $q=2^m$ support the complementary design of a $4$-$(q+1,7,\lambda_{1})$ design when $m \geqslant 5$ is odd and they support the complementary design of a $3$-$(q+1,7,\lambda_{2})$ design when $m \geqslant 4$ is even where
      $$\lambda_{1}=\frac{7(q-5)(q-8)}{6};\ \lambda_{2}=\frac{7(q-4)(q-5)(q-10)}{24}.$$
\item The codewords of weight $q-4$ in the trace code $\Tr_{q^{2}/q}(\mathcal{C}_{\{1,2,3\}})$ for $q=2^{2s+1}$ support the supplementary design of $(U_{q+1},\mathcal{B}_{\sigma_{5,3},q+1}^{b})$ which is isomorphic to the $4$-$(q+1,5,5)$ design constructed by Alltop \cite{1969An}.
\item We produced infinite families of simple $t$-designs with new parameters (in comparison with \cite{2006Handbook1} and \cite{Tang2020An}), which are summarized in Table 4 (complementary designs and supplementary designs not included).
\begin{table}[h]
\centering
\begin{tabular}{c|c|c|c}
    \hline
\text{Block sets} & \text{Designs} &\text{Conditions} & \text{Ref.}\\
   \hline
$\mathcal{B}_{\sigma_{5,3},q+1}^{\overline{b}}$  & $3$-$(q+1,5,\frac{q^2-10q+26}{2})$  &  $q=2^{2s}$ &Theorem  \ref{t3.5}\\
$\mathcal{B}_{\sigma_{7,3},q+1}^{u}$  & $4$-$(q+1,7,\frac{7(q-5)(q-8)}{6})$  &  $q=2^{2s+1}$ &Theorem \ref{t3.7} \\
$\mathcal{B}_{\sigma_{7,3},q+1}^{u}$  & $3$-$(q+1,7,\frac{7(q-4)(q-5)(q-10)}{24})$  &  $q=2^{2s}$ &Theorem \ref{t3.8} \\
$\mathcal{B}_{\sigma_{7,3},q+1}^{0}$  & $3$-$(q+1,7,\frac{7(q-4)(q-5)}{4})$  &  $q=2^{2s}$ &Theorem \ref{t3.2222} \\
    \hline
  \end{tabular}
\caption{$t$-Designs with new parameters}
\end{table}
\end{itemize}

An interesting open problem is to construct infinite families of $t$-designs with new parameters from elementary symmetric polynomials $\sigma_{k,l}$ and their variants, such as $(k,l)=(8,3)$, $(9,3)$ and $(10,3)$. Another open problem is whether there exist  linear codes holding $t$-designs isomorphic to those produced from ESPs and their variants.

\end{document}